\documentclass[twoside,12pt]{article}
\usepackage{epsfig}

\newcommand{\be}{\begin{equation}}
\newcommand{\ee}{\end{equation}}
\newcommand{\bea}{\begin{eqnarray}}
\newcommand{\eea}{\end{eqnarray}}

\usepackage{amsmath}
\usepackage{amsfonts}
\usepackage{amssymb}
\usepackage{amsxtra}
\begin{document}
\title{Clifford odd and even objects in even and odd dimensional spaces 
describing internal spaces of fermion and boson fields }
\author{Norma Susana Manko\v c Bor\v stnik\\ \\Department of Physics, University of 
Ljubljana\\
SI-1000 Ljubljana, Slovenia,\\ \\norma.mankoc@fmf.uni-lj.si\\}

\maketitle


\begin{abstract}
In a long series of works, it has been demonstrated, that the
{\it spin-charge-family} theory offers the explanation for all in the 
{\it standard model} assumed properties of the second quantized fermion 
and boson fields, offering several predictions as well as explanations for 
several of the observed phenomena.  The theory assumes a simple starting 
action in even dimensional spaces with $d \ge (13 +1)$ with massless 
fermions interacting with gravity only. The internal spaces of fermion and 
boson fields are described by the Clifford odd and even objects, respectively.  
This contribution discusses the properties of the fermion and boson fields in odd 
dimensional spaces, $d=(2n +1)$, with the internal spaces of fermion and 
boson fields described again by the Clifford odd and even objects, respectively, 
pointing out that their properties differ essentially from the properties in even 
dimensional spaces, resembling the ghost needed when looking for final 
solutions with Feynman diagrams.
\end{abstract}
\section{introduction}
\label{introduction}
%

30 years ago, I recognized that there are two kinds of Clifford algebra objects, 
$\gamma^a$'s and $\tilde{\gamma}^a$'s~\cite{norma93,norma92,norma95}, 
originating in the Grassmann algebra. The Clifford and the Grassmann algebras 
can be used to describe the internal 
space of fermions in even dimensional spaces: The superposition of odd products of 
either $\gamma^a$'s or $\tilde{\gamma}^a$'s, anti-commute, fulfilling on the 
vacuum states the anti-commutation relations~\cite{2020PartIPartII} of the second 
quantization postulates for fermion fields~\cite{Dirac,BetheJackiw}. The superposition 
of odd products of either $\gamma^a$'s or $\tilde{\gamma}^a$'s, appear in 
irreducible representations~\cite{nh2021RPPNP}.

Only one kind of fermions has been observed so far, appearing in several families. If
we use one of the two kinds of the Clifford algebra objects, say $\gamma^a$'s, to 
describe the internal space of fermions, and the second kind of the Clifford algebra 
objects, $\tilde{\gamma}^a$'s, to describe the family quantum numbers of each 
of the irreducible representation determined by $\gamma^a$'s, we are left with 
one kind of fermions~\cite{JMP2013,nh03}, Sect.(3.2.3) of~\cite{nh2021RPPNP}. 

In any even dimensional space there are $2^{\frac{d}{2}-1}$ of the Clifford odd
``basis vectors'', appearing in $2^{\frac{d}{2}-1}$ families.  They are the superposition 
of odd products $\gamma^a$'s. All the members of any family are orthogonal to
all the members of the same and all the other families. Their Hermitian 
conjugated partners appear in a separate group, again with $2^{\frac{d}{2}-1}$ 
members in $2^{\frac{d}{2}-1}$ families.

The Clifford odd ``basis vectors'' have in even dimensional spaces only left or only 
right handedness, depending on the definition ($ \Gamma= \prod_a^d (\sqrt{\eta^{aa}} \gamma^a)  \cdot  (i)^{\frac{d}{2}}$).

In any even dimensional space there are two groups of $2^{\frac{d}{2}-1}\times $
$2^{\frac{d}{2}-1}$ of the Clifford even ``basis vectors'', which are the superposition 
of even products $\gamma^a$'s. The family quantum number has no meaning  for
the Clifford even ``basis vectors''. The members of one group are orthogonal to the
members of another group. The members of any of the two  groups  of the Clifford 
even ``basis vectors'' have their Hermitian conjugated partners within the same 
group~\cite{n2021SQ,n2022epjc}.

The superposition of even products of $\gamma^a$'s (or $\tilde{\gamma}^a$'s), 
commute, fulfilling the commutation relations~\cite{2020PartIPartII} of the second 
quantization postulates for boson fields~\cite{Dirac,BetheJackiw,Weinberg}.

The Clifford even ``basis vectors'' have properties of the gauge fields of the 
corresponding Clifford odd ``basis vectors'', what becomes transparent after the 
algebraic multiplication, $*_A$, of the Clifford even  ``basis vectors''  on the 
Clifford odd ``basis vectors'' and opposite, as well as of the Clifford even 
``basis vectors'' among themselves~\cite{n2021SQ,n2022epjc}. 

Algebraic multiplication is distributive and associative.

The properties of the Clifford odd and the Clifford even ``basis vectors'' in even 
dimensional spaces is shortly overviewed in Sect.~\ref{evend}, showing that
the Clifford odd ``basis vectors'', applying on the appropriate vacuum states, 
manifest the postulates of the second quantized fermion fields, while the 
Clifford even ``basis vectors'' manifest the postulates for their gauge fields,
the second quantized boson fields.

The properties of the fermion and boson fields in odd dimensional spaces differ  
drastically  from the properties of the  fermion and boson fields 
in even dimensional spaces: The Clifford odd ``basis vectors'' do not manifest 
the properties of the second quantized fermion fields in even dimensional spaces. 
Although anti-commuting, they instead manifest properties of the  Clifford even 
``basis vectors'' in even dimensional spaces. And the Clifford even ``basis 
vectors'' do not manifest the properties of the second quantized boson fields in 
even dimensional spaces. Although commuting, they instead manifest properties 
of the  Clifford odd ``basis vectors'' in even dimensional spaces.

In addition, since the operator of handedness has in odd dimensional spaces 
the Clifford odd character 
($ \Gamma= \prod_a^d (\sqrt{\eta^{aa}} \gamma^a) \cdot  (i)^{\frac{d-1}{2}}$), 
it transforms the  Clifford odd ``basis vectors'' into the Clifford even ``basis vectors''~\cite{n2022odd}. 
 The eigenstates of the operator of handedness are in odd  dimensional spaces
 correspondingly the superposition of the Clifford odd and the Clifford even ``basis 
 vectors''. 
 
 The properties of the Clifford odd and the Clifford even ''basis vectors'' in odd 
dimensional spaces are discussed in Sect.~\ref{ODDd}.


In $d=(13 +1)$ dimensional space the Clifford odd ``basis vectors'', if analysed from 
the point of view of the subgroups of the {\it standard model} groups, offer the 
description of the internal spaces of all the so far observed quarks and leptons and 
antiquarks and antileptons as assumed by the  {\it standard model} before the 
electroweak phase transition, including in addition the right handed neutrinos and left 
handed antineutrions. Quarks and antiquarks and leptons and antileptons appear as 
sixty-four (64) members in two times four families.  

The corresponding Clifford even ``basis vectors'' offer the description of the internal 
spaces of the corresponding vector and scalar gauge fields~\cite{nh2021RPPNP,%
n2021SQ,n2022epjc,nBled2022}. 

The {\it spin-charge-family} theory, describing the internal spaces of fermion and 
boson fields by using the Clifford odd and even algebras in $d=(13+1)$-dimensional 
space, offers not only the explanation for the postulates of the second quantized 
fermion and boson fields, and the explanation for all the {\it standard model}
assumptions, but also for several observed phenomena, making several 
predictions~\cite{nh2021RPPNP}. The theory is built on the simple starting starting 
action in which fermion interacts with the gravitational fields only
\begin{eqnarray}
{\cal A}\,  &=& \int \; d^dx \; E\;\frac{1}{2}\, (\bar{\psi} \, \gamma^a p_{0a} \psi) 
+ h.c. +
\nonumber\\  
               & & \int \; d^dx \; E\; (\alpha \,R + \tilde{\alpha} \, \tilde{R})\,,
\nonumber\\
               p_{0a } &=& f^{\alpha}{}_a p_{0\alpha} + \frac{1}{2E}\, \{ p_{\alpha},
E f^{\alpha}{}_a\}_- \,,\nonumber\\
          p_{0\alpha} &=&  p_{\alpha}  - \frac{1}{2}  S^{ab} \omega_{ab \alpha} - 
                    \frac{1}{2}  \tilde{S}^{ab}   \tilde{\omega}_{ab \alpha} \,,
                    \nonumber\\                    
R &=&  \frac{1}{2} \, \{ f^{\alpha [ a} f^{\beta b ]} \;(\omega_{a b \alpha, \beta} 
- \omega_{c a \alpha}\,\omega^{c}{}_{b \beta}) \} + h.c. \,, \nonumber \\
\tilde{R}  &=&  \frac{1}{2} \, \{ f^{\alpha [ a} f^{\beta b ]} 
\;(\tilde{\omega}_{a b \alpha,\beta} - \tilde{\omega}_{c a \alpha} \,
\tilde{\omega}^{c}{}_{b \beta})\} + h.c.\,.               
\label{wholeaction}
\end{eqnarray}
Here~\footnote{$f^{\alpha}{}_{a}$ are inverted vielbeins to 
$e^{a}{}_{\alpha}$ with the properties $e^a{}_{\alpha} f^{\alpha}{\!}_b = 
\delta^a{\!}_b,\; e^a{\!}_{\alpha} f^{\beta}{\!}_a = \delta^{\beta}_{\alpha} $, 
$ E = \det(e^a{\!}_{\alpha}) $.
Latin indices  
$a,b,..,m,n,..,s,t,..$ denote a tangent space (a flat index),
while Greek indices $\alpha, \beta,..,\mu, \nu,.. \sigma,\tau, ..$ denote an Einstein 
index (a curved index). Letters  from the beginning of both the alphabets
indicate a general index ($a,b,c,..$   and $\alpha, \beta, \gamma,.. $ ), 
from the middle of both the alphabets   
the observed dimensions $0,1,2,3$ ($m,n,..$ and $\mu,\nu,..$), indexes from 
the bottom of the alphabets
indicate the compactified dimensions ($s,t,..$ and $\sigma,\tau,..$). 
We assume the signature $\eta^{ab} =
diag\{1,-1,-1,\cdots,-1\}$.} 
$f^{\alpha [a} f^{\beta b]}= f^{\alpha a} f^{\beta b} - f^{\alpha b} f^{\beta a}$.


I demonstrate in this paper that in odd dimensional spaces the Clifford odd and 
the Clifford even objects have drastically different properties than in even dimensional
spaces, offering the explanation for postulated ghost fields appearing in several theories for 
taking care of the singular contributions in evaluating Feynman graphs.

In Sect.~\ref{generalproperties} appropriate definition of the eigenstates of the 
Cartan subalgebra members are presented for even dimensional spaces, and extended to
odd dimensional spaces.  

In Subsect.~\ref{evend}  the internal spaces described by the Clifford odd and the 
Clifford even ''basis vectors'' for fermion and boson fields in even dimensional spaces
are presented.

In Subsect.~\ref{ODDd} the internal spaces of fermion and boson fields in odd 
dimensional spaces are presented.

In Sect.~\ref{specialcases},  the internal spaces for fermion and boson fields in even 
and odd dimensional spaces for simple cases are discussed: In 
Subsect.~\ref{2n} for the choices $d=(1+1)$, $d=(3+1)$ and in 
Subsect.~\ref{ODDdspecial} for 
$d=(2+1)$ and $d=(4+1)$.

In Refs.~\cite{nh2000,nh2002,nhl2002} from 20 years ago the 
authors discuss the question of $q$ time  and $d-q$ dimensions in odd and 
even dimensional spaces for any $q$. Using the requirements that the inner product of 
two fermions is unitary and invariant under Lorentz transformations the authors conclude
that odd dimensional spaces are not appropriate due to the existence of fermions of both handedness and correspondingly not mass protected. 
The recognition of this paper might further 
clarify the ``effective'' choice of Nature for one time and three space dimensions.

In Sect.~\ref{discussion}, the main idea of this note is overviewed.

In App.{A},  some helpful relations of the Clifford algebra can be found.


%
\section{Eigenstates of Cartan subalgebra members of  Lorentz algebra for 
Clifford odd and Clifford even ``basis vectors''}
\label{generalproperties}

In this section, the properties of the two kinds of Clifford algebra objects, 
$\gamma^a$'s and $\tilde{\gamma}^a$'s, are shortly repeated following several 
papers~\cite{norma93,norma92,JMP2013,2020PartIPartII,n2021SQ,nBled2022,%
n2022epjc,nIARD2022}, in particular the reference~(\cite{nh2021RPPNP}, and the
references therein).  

The two kinds of Clifford algebra objects, $\gamma^a$ and $\tilde{\gamma}^a$,
each offering $2^d$ superposition of products of either $\gamma^a$ or
 $\tilde{\gamma}^a$, fulfil the relation~\cite{norma93,nh02,nh03}
\begin{eqnarray}
\label{gammatildeantiher}
\{\gamma^{a}, \gamma^{b}\}_{+}&=&2 \eta^{a b}= \{\tilde{\gamma}^{a}, 
\tilde{\gamma}^{b}\}_{+}\,, \nonumber\\
\{\gamma^{a}, \tilde{\gamma}^{b}\}_{+}&=&0\,,\quad
 (a,b)=(0,1,2,3,5,\cdots,d)\,, \nonumber\\
(\gamma^{a})^{\dagger} &=& \eta^{aa}\, \gamma^{a}\, , \quad 
(\tilde{\gamma}^{a})^{\dagger} =  \eta^{a a}\, \tilde{\gamma}^{a}\,.
\end{eqnarray}
Each of these two kinds of the Clifford algebra objects  could be used to describe the
internal spaces of fermion and boson fields. 

We can reduce the two possibilities to only one by deciding to describe 
the internal spaces of fermion and boson fields with the  superposition of the Clifford 
odd  (for fermion fields) and the Clifford even (for boson fields) products of 
$\gamma^a$'s,  while using  $\tilde{\gamma}^a$'s to equip the irreducible 
representations of the Lorentz group in the internal space of fermions with the 
family quantum numbers by assuming
\begin{eqnarray}
\{\tilde{\gamma}^a B &=&(-)^B\, i \, B \gamma^a\}\, |\psi_{oc}>\,,
\label{tildegammareduced}
\end{eqnarray}
with $(-)^B = -1$, if $B$ is (a function of) an odd product of $\gamma^a$'s,
 otherwise $(-)^B = 1$~\cite{nh03}, $|\psi_{oc}>$ is defined in 
Eq.~(\ref{vaccliffodd}). 
It is proven in~\cite{nh2021RPPNP}~(App.I, Statement 3, 3.a, 3.b) that all the 
relations of Eq.~(\ref{gammatildeantiher}) remain valid also after the assumption of 
Eq.~(\ref{tildegammareduced}).

The ``basis vectors''  describing internal spaces of fermion and boson fields are 
chosen to be eigenstates of all the Cartan subalgebra members. There are 
$\frac{d}{2}$ commuting operators of the Lorentz algebra in the even 
dimensional spaces, Eq.~(\ref{cartancliffevend}), and  $\frac{d-1}{2}$ in odd 
dimensional spaces, Eq.~(\ref{cartancliffODDd}).

If $S^{ab}$, $a\ne b$, (or $\tilde{S}^{ab}$ or ${\cal {\bf S}}^{ab} = 
S^{ab} +\tilde{S}^{ab}$) are members of the Cartan subalgebra group of the 
Lorentz algebra in the internal space of fermion and boson fields, then it is not 
difficult to find the eigenstate of each of the members just by taking into account 
relations of Eq.~(\ref{gammatildeantiher}: $S^{ab} \,\frac{1}{2} (\gamma^a + \frac{\eta^{aa}}{ik} \gamma^b) = \frac{k}{2}  
\,\frac{1}{2} (\gamma^a + \frac{\eta^{aa}}{ik} \gamma^b)$ and 
$S^{ab}\, \frac{1}{2} (1 +  \frac{i}{k}  \gamma^a \gamma^b) = \frac{k}{2}  \,
 \frac{1}{2} (1 +  \frac{i}{k}  \gamma^a \gamma^b)$, with  
 $k^2=\eta^{aa} \eta^{bb}$. The first eigenstate is nilpotent, 
 $(\frac{1}{2} (\gamma^a + \frac{\eta^{aa}}{ik} \gamma^b))^2=0$ and the second eigenstate is projector $(\frac{1}{2} (1 +  \frac{i}{k}  \gamma^a \gamma^b))^2 =
 \frac{1}{2} (1 +  \frac{i}{k}  \gamma^a \gamma^b)$.

 Let us introduce the graphic notation, following Ref.~\cite{n2022epjc,nh02,nh03}.
\begin{eqnarray}
\label{graphic}
\stackrel{ab}{(k)}:&=& 
\frac{1}{2}(\gamma^a + \frac{\eta^{aa}}{ik} \gamma^b)\,,\quad 
\stackrel{ab}{[k]}:=\frac{1}{2}(1+ \frac{i}{k} \gamma^a \gamma^b)\,,\nonumber\\
\stackrel{ab}{\tilde{(k)}}:&=& 
\frac{1}{2}(\tilde{\gamma}^a + \frac{\eta^{aa}}{ik} \tilde{\gamma}^b)\,,\quad 
\stackrel{ab}{\tilde{[k]}}:
\frac{1}{2}(1+ \frac{i}{k} \tilde{\gamma}^a \tilde{\gamma}^b)\,,\nonumber\\
(\stackrel{ab}{(k)})^{\dagger}&=& \stackrel{ab}{(-k)}\,, \quad
(\stackrel{ab}{(k)})^{2}= 0\,, \quad (\stackrel{ab}{[k]})^{\dagger}= \stackrel{ab}{[k]}\,,
\quad (\stackrel{ab}{[k]})^{2}= \stackrel{ab}{[k]}\,.
\end{eqnarray}
After taking into account Eq.~(\ref{gammatildeantiher}) the relations  follow
\begin{eqnarray}
\label{graphicfollow}
\gamma^a \stackrel{ab}{(k)}&=& \eta^{aa}\stackrel{ab}{[-k]},\; \quad
\gamma^b \stackrel{ab}{(k)}= -ik \stackrel{ab}{[-k]}, \; \quad 
\gamma^a \stackrel{ab}{[k]}= \stackrel{ab}{(-k)},\;\quad \;\;
\gamma^b \stackrel{ab}{[k]}= -ik \eta^{aa} \stackrel{ab}{(-k)}\,,\nonumber\\
\tilde{\gamma^a} \stackrel{ab}{(k)} &=& - i\eta^{aa}\stackrel{ab}{[k]},\quad
\tilde{\gamma^b} \stackrel{ab}{(k)} =  - k \stackrel{ab}{[k]}, \;\qquad  \,
\tilde{\gamma^a} \stackrel{ab}{[k]} =  \;\;i\stackrel{ab}{(k)},\; \quad
\tilde{\gamma^b} \stackrel{ab}{[k]} =  -k \eta^{aa} \stackrel{ab}{(k)}\,, 
\end{eqnarray}

More relations can be found in App.~\ref{A}.

%
%

%
\subsection{Properties of Clifford odd and Clifford even ``basis vectors'' in even
dimensional spaces}
\label{evend}

In each even dimensional space there are $2^{\frac{d}{2}-1}$ members of the Clifford
odd ``basis vectors'' appearing $2^{\frac{d}{2}-1}$ families, and the same number
of $2^{\frac{d}{2}-1}$ their Hermitian conjugated partners  appearing in 
$2^{\frac{d}{2}-1}$ families.

There are two orthogonal groups of the Clifford even ``basis vectors''. The members
of each group have their Hermitian conjugated partners within the same group.

\vspace{3mm}

{\bf Clifford odd  ``basis vectors''}

\vspace{2mm}

We find the Clifford odd ``basis vectors'', describing the internal space of fermion 
fields, as products of odd numbers of nilpotents and the rest of projectors, if each 
nilpotent and each projector is the eigenstate of one of the Cartan subalgebra 
members.

Let us call the Clifford odd ''basis vectors''  $\hat{b}^{m \dagger}_f$, if this is the
$m^{th}$ member of the family $f$. 

Let us choose the first member  $\hat{b}^{1 \dagger}_1$, if 
$d=2(2n+1)$, as the product of nilpotents only. 
%
\begin{small}
\begin{eqnarray}
\label{allcartaneigenvecb}
&& d=2(2n+1)\, ,\nonumber\\
&& \hat{b}^{1 \dagger}_{1}=\stackrel{03}{(+i)}\stackrel{12}{(+)} \stackrel{56}{(+)}
\cdots \stackrel{d-1 \, d}{(+)}\,,\nonumber\\
&&\hat{b}^{2 \dagger}_{1} = \stackrel{03}{[-i]} \stackrel{12}{[-]} 
\stackrel{56}{(+)} \cdots \stackrel{d-1 \, d}{(+)}\,,\nonumber\\
&& \cdots\nonumber\\
&&\hat{b}^{2^{\frac{d}{2}-1} \dagger}_{1} = \stackrel{03}{[-i]} \stackrel{12}{[-]} 
\stackrel{56}{(+)} \dots \stackrel{d-3\,d-2}{[-]}\;\stackrel{d-1\,d}{[-]}\,, \nonumber\\
&& \cdots\,.
\end{eqnarray}
\end{small}
In the case that $d=4n, n=1,2,..,$
the first member must have one projector.  
\begin{small}
\begin{eqnarray}
\label{allcartaneigenvecb4n}
&& d=4n\, ,\nonumber\\
&& \hat{b}^{1 \dagger}_{1}=\stackrel{03}{(+i)}\stackrel{12}{(+)} \stackrel{56}{(+)}
\cdots \stackrel{d-1 \, d}{[+]}\,,
 \nonumber\\
&& \cdots\,.
\end{eqnarray}
\end{small}
All the rest members of the same family,
$2^{\frac{d}{2}-1}-1$, follow by the application of all possible $S^{ab}$ 
on  $\hat{b}^{1 \dagger}_1$, while all the rest $2^{\frac{d}{2}-1}-1$ families follow
by the application of all possible $\tilde{S}^{ab}$ on all the members of the starting 
family.
%


The Hermitian conjugated partners $(\hat{b}^{m \dagger}_f)^{\dagger}$  of the 
``basis vectors'' $\hat{b}^{m \dagger}_f$ follow from these 
$2^{\frac{d}{2}-1}\times 2^{\frac{d}{2}-1}$ ``basis vectors'' by replacing each 
nilpotent $\stackrel{ab}{(k)}$ with $\stackrel{ab}{(-k)}$.

Choosing the vacuum  state equal to
\begin{eqnarray}
\label{vaccliffodd0}
|\psi_{oc}>= \sum_{f=1}^{2^{\frac{d}{2}-1}}\,\hat{b}^{m}_{f}{}_{*_A}
\hat{b}^{m \dagger}_{f} \,|\,1\,>\,,
\end{eqnarray}
for one of the members $m$, anyone, of the odd irreducible representation $f$,
with $|\,1\,>$, which is the vacuum without any structure --- the identity ---
it follows that $\hat{b}^{m}_{f}{} |\psi_{oc}>=0$. 

Each  Clifford odd ``basis vector'' carries the family quantum number, and so 
does its Hermitian conjugated partner.
One correspondingly finds that the ``basis vectors'' and their Hermitian conjugated
partners fulfil the postulates for the second quantized fermion fields.
%
%
\begin{eqnarray}
\label{Dirac}
\hat{b}^{m}_{f} {}_{*_{A}}|\psi_{oc}>&=& 0.\, |\psi_{oc}>\,,\nonumber\\
\hat{b}^{m \dagger}_{f}{}_{*_{A}}|\psi_{oc}>&=&  |\psi^m_{f}>\,,\nonumber\\
\{\hat{b}^{m}_{f}, \hat{b}^{m'}_{f `}\}_{*_{A}+}|\psi_{oc}>&=&
 0.\,|\psi_{oc}>\,, \nonumber\\
\{\hat{b}^{m \dagger}_{f}, \hat{b}^{m' \dagger}_{f  `}\}_{*_{A}+}|\psi_{oc}>
&=& 0. \,|\psi_{oc}>\,,\nonumber\\
\{\hat{b}^{m}_{f}, \hat{b}^{m' \dagger}_{f `}\}_{*_{A}+}|\psi_{oc}>
&=& \delta^{m m'}_{f f `}|\psi_{oc}>\,,
\end{eqnarray}
where  $*_{A}$ represents the algebraic multiplication of 
$\hat{b}^{m \dagger}_{f}$  and $ \hat{b}^{m'}_{f'} $  among themselves and  
with the vacuum state  $|\psi_{oc}>$ of Eq.(\ref{vaccliffodd0}). Eq.~(\ref{Dirac}) 
follows by taking into account Eq.~(\ref{gammatildeantiher}). 


These ``basis vectors'' are not yet the representatives of the creation and annihilation
operators:  They must be tensor, $*_{T}$, products of the ``basis vectors" and the basis in ordinary momentum or coordinate space~\cite{nh2021RPPNP}~\footnote{
In even dimensional spaces with $d=4n$, one proceeds  as we did in 
$d=2(2n+1)$ dimensional case after taking into account the requirement that 
the odd number of nilpotents forms the anti-commuting ``basis vectors'' describing 
the internal space of fermions:
The starting  ``basis vector'' $\hat{b}^{1 \dagger}_{1}$ must have one projector,
while all the rest are nilpotents. $S^{ab}$'s then generate all the members of one family,
while $\tilde{S}^{ab}$'s generate all the families. The ``basis vectors'' and their 
Hermitian conjugated partners fulfil on the vacuum state, Eq.~(\ref{vaccliffodd}), 
the anti-commuting postulates of Eq.~(\ref{Dirac}).}.

\vspace{3mm}

{\bf Clifford even ``basis vectors''}

\vspace{2mm}

We can find the Clifford even ``basis vectors'' describing the internal space of the boson 
fields as products of even numbers of nilpotents and the rest of projectors if each 
nilpotent and each projector is the eigenstate of one of the Cartan subalgebra 
members. 

Let us call the Clifford even   ``basis vectors''  
${}^{i}{\bf {\cal A}}^{m \dagger}_{f}, i=I,II$. 
There are namely two groups of Clifford even basis vectors''. Each group has 
$2^{\frac{d}{2}-1}\times 2^{\frac{d}{2}-1}$ members. 

Let us choose the starting Clifford even ``basis vector'',  
${}^{i=I}{\bf {\cal A}}^{1 \dagger}_{1}$, to be the product of projectors  
$\stackrel{ab}{[k]}$, with
$k=i$ for $S^{03}$,  and $k=1$ for the rest $2^{\frac{d}{2}-1}-1 $ members of the 
Cartan subalgebra.
\begin{small}
\begin{eqnarray}
\label{allcartaneigenvecevenI} 
{}^I\hat{{\cal A}}^{1 \dagger}_{1}=\stackrel{03}{[+i]}\stackrel{12}{[+]}\cdots 
\stackrel{d-1 \, d}{[+]}\,.
\end{eqnarray}
\end{small}
The starting Clifford even ``basis vector''  of the second group 
${}^{i=II}{\bf {\cal A}}^{1 \dagger}_{1}$  can again be the product  of  projectors 
only, but in this case with $\stackrel{03}{[-i]}$ instead of $\stackrel{03}{[+i]}$ and 
for all the rest $2^{\frac{d}{2}-1}-1 $ members of the Cartan subalgebra with $k=+1$. 
(This starting member can not be obtained from 
${}^{I}{\bf {\cal A}}^{1 \dagger}_{1}$ by the application of $S^{ab}$'s or 
$\tilde{S}^{ab}$'s, since  these operators always change the eigenvalues of  two 
Cartan subalgebra members.)
\begin{small}
\begin{eqnarray}
\label{allcartaneigenvecevenII} 
{}^{II}\hat{{\cal A}}^{1 \dagger}_{1}=\stackrel{03}{[-i]}\stackrel{12}{[+]}\cdots 
\stackrel{d-1 \, d}{[+]}\,.
\end{eqnarray}
\end{small}

 The rest of the members of each group follow from the 
starting member by the application of either $S^{ab}$'s or $\tilde{S}^{ab}$'s.

Since  $ S^{0 1}$ transforms  $\stackrel{03}{[+i]} \stackrel{12}{[+]}$ into
$\stackrel{03}{(-i)} \stackrel{12}{(-1)}$, while $\tilde{S}^{0 1}$ transforms  
$\stackrel{03}{[+i]} \stackrel{12}{[+]}$ into $\stackrel{03}{(+i)} \stackrel{ab}{(+)}$,
we immediately see that the Clifford even ``basis vector''  have the Hermitian conjugated
partners within the same group of $2^{\frac{d}{2}-1}\times 2^{\frac{d}{2}-1}$ 
members.

\vspace{3mm}

{\bf Clifford even ``basis vectors'' applying on Clifford odd ``basis vectors.}

\vspace{2mm}

Let us apply ${}^{I}{\bf {\cal A}}^{1 \dagger}_{1}$, which is made of projectors  
$\stackrel{ab}{[k]}$ only, with $k=i$ for $S^{03}$,  and $k=1$ for the rest members 
of the Cartan subalgebra, on $\hat{b}^{1 \dagger}_{1}$, which is the product of 
nilpotents only, with eigenvalue of $S^{03}$ equal $k=i$ and of the rest of Cartan 
subalgebra members equal to $k=1$.  

Taking into account Eqs.~(\ref{graficcliff1}, \ref{graficfollow1}) one sees that
this application, ${}^{I}{\bf {\cal A}}^{1 \dagger}_{1} *_{A} 
\hat{b}^{1 \dagger}_{1}$, leaves $ \hat{b}^{1 \dagger}_{1}$ unchanged. 
When applying ${}^{I}{\bf {\cal A}}^{2 \dagger}_{1}$, with the first two 
projectors transformed into two nilpotents, $\stackrel{03}{(-i)} \stackrel{12}{(-1)}$, 
and all the rest remain the same, we see that this application transforms 
$\hat{b}^{1 \dagger}_{1} $ into $\hat{b}^{2 \dagger}_{1}$ 
($=\stackrel{03}{[-i]} \stackrel{12}{[-1]} \stackrel{56}{(+)} \stackrel{78}{(+)}....$ 
(all the rest remains the same). The application of ${}^{I}{\bf {\cal A}}^{2 \dagger}_{1}$ on
$ \hat{b}^{1 \dagger}_{1}$ obviously changes the eigenvalues of $S^{03}$ and 
of $S^{12}$ of  $ \hat{b}^{1 \dagger}_{1}$ for integer values, 
$-i$ and $-1$, respectively. 

We conclude: The algebraic application, $*_{A}$, of the Clifford even ''basis vectors''
on the Clifford odd ''basis vectors'', describing the internal space of fermion fields, 
 change their eigenvalues of the Cartan subalgebra members for $0$ or
for integer values, $\pm i$, or $\pm 1$, leading to
\begin{eqnarray}
\label{Abgen}
{}^{I}{\hat{\cal A}}^{m \dagger}_{f `} \,*_A \, \hat{b}^{m' \dagger }_{f}
\rightarrow \left \{ \begin{array} {r} \hat{b }^{m \dagger}_{f }\,, \\
{\rm or \,zero}\,.
\end{array} \right.
\end{eqnarray}

\vspace{3mm}

{\bf Clifford even ``basis vectors'' applying on Clifford even ``basis vectors''}

\vspace{2mm}

It is not difficult to see, by taking into account Eqs.~(\ref{graficcliff1}, \ref{graficfollow1}), 
that the algebraic applications of ${}^{I}{\bf {\cal A}}^{f \dagger}_{1} *_{A} 
{}^{II}{\bf {\cal A}}^{m' \dagger}_{f `}=0={}^{II}{\bf {\cal A}}^{m' \dagger}_{f `}
*_{A} {}^{I}{\bf {\cal A}}^{m \dagger}_{f }$, for all ($m, m', f, f `$).

The algebraic application, $*_{A}$, of ${}^{i}{\bf {\cal A}}^{m\dagger}_{f} *_{A} 
{}^{i}{\bf {\cal A}}^{m' \dagger}_{f `}$ within each of the two groups give in 
general non zero contribution, demonstrating the properties of the internal spaces of
the gauge fields to the corresponding fermion fields, the internal space of which
are described by the Clifford odd ``basis vectors''.  

In each of the two groups, there are  $2^{\frac{d}{2}-1}$ members, which are products 
of projectors only. They are self adjoint and have the eigenvalues of all the Cartan 
subalgebra members equal zero: ${\bf {\cal S}}^{ab}= S^{ab} + \tilde{S}^{ab}$. 

All the rest ${}^{i}{\bf {\cal A}}^{m \dagger}_{f}$ (there are $2^{\frac{d}{2}-1}
\times (2^{\frac{d}{2}-1}-1)$ members)  appear in pairs; Hermitian conjugated 
to each other. Their mutual algebraic products form one of $2^{\frac{d}{2}-1}$
self-adjoint members. 

The algebraic multiplication of the Clifford even ``basis vectors'' on the Clifford even
``basis vectors'' lead to
\begin{eqnarray}
\label{AAgen}
{}^{i}{\hat{\cal A}}^{m \dagger}_{f} \,*_A\, {}^{i}{\hat{\cal A}}^{m' \dagger}_{f `}
\rightarrow \left \{ \begin{array} {r} {}^{i}{\hat{\cal A}}^{m \dagger}_{f `}\,, 
\\
{\rm or \,zero}\,.
\end{array} \right.
i=(I,II)\,.
\end{eqnarray}

The reader can find in Ref.~\cite{nBled2022,n2022epjc} the Clifford odd and the Clifford even 
''basis vectors'' in the case that the dimension of the space is $d=(5+1)$, describing 
the internal space of fermion and boson fields, respectively, illustrated by figures.

\subsection{Properties of the Clifford odd and Clifford even ''basis vectors'' in odd
dimensional spaces}
\label{ODDd}
%
%

In this Subsect.~\ref{ODDd} the Clifford odd and Clifford even ``basis vectors'' in 
odd dimensional spaces~\cite{n2021SQ,n2022epjc} are discussed.

While in even dimensional spaces the Clifford odd ``basis vectors'' fulfil the postulates for 
the second quantized fermion fields, Eq.~(\ref{Dirac}), and Clifford even ''basis vectors'' 
have all the properties of the internal spaces of their corresponding gauge fields, Eqs.~(\ref{Abgen}, \ref{AAgen}), the Clifford odd and even ''basis vectors'' have 
in odd dimensional spaces unusual properties resembling properties of the internal 
spaces of the Faddeev-Popov ghosts, as we shall see in what follows.

Looking in $d=(2n+1)$dimensional cases, $n=1,2,\dots$,  for the Clifford odd and 
Clifford even ``basis vectors'' in $2n$-dimensional part of space we find half of the 
``basis vectors'' with properties presented in Eqs.~(\ref{allcartaneigenvecb}, 
\ref{allcartaneigenvecb4n}, \ref{allcartaneigenvecevenI}).  
In Eqs.~(\ref{allcartaneigenvecbdgen}, \ref{allcartaneigenvecAdgen}) they are 
presented on the left hand side.

The rest of  the ``basis vectors'' follow applying $S^{0 \,2n+1}$  on the 
obtained half of  the  Clifford odd and the Clifford even ``basis vectors''. Since 
$S^{0 \,2n+1}$ are Clifford even operators;  they do not change oddness or 
evenness of the ``basis vectors''. 

One finds for the Clifford odd ``basis vectors'' correspondingly the  additional 
$2^{\frac{d-1}{2}-1}$ members, appearing in $2^{\frac{d-1}{2}-1}$ families 
and the same number of their Hermitian conjugated partners on the right hand 
side of Eq.~(\ref{allcartaneigenvecbdgen}).  
\begin{small}
\begin{eqnarray}
\label{allcartaneigenvecbdgen}
 d=&&2(2n+1)+1\, \nonumber\\
 \hat{b}^{1 \dagger}_{1}=\stackrel{03}{(+i)}\stackrel{12}{(+)} \stackrel{56}{(+)}
\cdots \stackrel{d-2 \, d-1}{(+)} \,,\quad 
&& \hat{b}^{1 \dagger}_{2^{\frac{d-1}{2}-1}+1}=\stackrel{03}{[-i]}\stackrel{12}{(+)} \stackrel{56}{(+)}
\cdots \stackrel{d-2 \, d-1}{(+)} \gamma^{d}\,,\nonumber\\
\hat{b}^{2 \dagger}_{1} = \stackrel{03}{[-i]} \stackrel{12}{[-]} 
\stackrel{56}{(+)} \cdots \stackrel{d-2 \, d-1}{(+)}\,, \quad
&&\hat{b}^{2 \dagger}_{2^{\frac{d-1}{2}-1}+1} = \stackrel{03}{(+i)} \stackrel{12}{[-]} 
\stackrel{56}{(+)} \cdots \stackrel{d-2 \, d-1}{(+)} \gamma^{d}\,,\nonumber\\
\cdots \quad 
&& \cdots\nonumber\\
\hat{b}^{2^{\frac{d-1}{2}-1} \dagger}_{1} = \stackrel{03}{[-i]} \stackrel{12}{[-]} 
\stackrel{56}{(+)} \dots \stackrel{d-2\,d-1}{[-]}\;, \quad
&&\hat{b}^{2^{\frac{d-1}{2}-1} \dagger}_{2^{{d-1}{2}-1}+1} = \stackrel{03}{(+i)} \stackrel{12}{[-]} 
\stackrel{56}{(+)} \dots \stackrel{d-2\,d-1}{[-]}\; \gamma^{d}\,, \nonumber\\
\cdots \quad
&& \cdots \,.
\end{eqnarray}
\end{small}
The right handed half of ``basis vectors''  follows from the left handed ``basis vectors''
or from their Hermitian conjugated partners by the application of $S^{0d}$ on the 
left handed part. The application of $\tilde{S}^{0d}$ on the left handed part of the 
``basis vectors'' generates the whole set of $2^{d-2}$ members of the Clifford odd 
''basis vectors'' from the right hand side~\footnote{
The application of $S^{0d}$ and $\tilde{S}^{0d}$ on the left hand side part of the 
Hermitian conjugated group to the Clifford odd ''basis vectors'' generate the same 
$2^{d-2}$ Clifford odd ``basis vectors'' as the $S^{0\,d}$ and $\tilde{S}^{0\,d}$ 
when applying on the left hand side ``basis vectors''. Correspondingly we now have 
twice $2^{d-2}$ Clifford odd eigenstates of the $\frac{d-1}{2}$ Cartan subalgebra 
members.}.

When applying on  the Clifford even ``basis vectors''  appearing on the left hand 
side of  Eq.~(\ref{allcartaneigenvecAdgen}) the operators $S^{0 \,2n+1}$ 
the additional two groups of $2^{\frac{d-1}{2}-1}\times$ $2^{\frac{d-1}{2}-1}$ 
``basis vectors'' follow, presented in Eq.~(\ref{allcartaneigenvecAdgen})  on the 
right hand side.

The $2^{d-2}$ Clifford odd objects presented on the right hand side of 
Eq.~(\ref{allcartaneigenvecbdgen}), and for the special cases of 
Eqs.~(\ref{allcartaneigenvecbd2+1}, \ref{allcartaneigenvecbd4+1}), although 
they are the superposition of  the Clifford odd products of $\gamma^a$'s, do 
not manifest properties of ``basis vectors'' and their Hermitian conjugated partners, 
presented on the left hand side of Eq.~(\ref{allcartaneigenvecbdgen}),  and for the 
special cases of Eqs.~( \ref{allcartaneigenvecbd2+1}, \ref{allcartaneigenvecbd4+1}).

The eigenstates appearing on the right hand side of Eq.~(\ref{allcartaneigenvecbdgen}) 
can be divided into two groups which are orthogonal to each other,  having their 
Hermitian conjugated partners within the same group or are self adjoint. Although 
they are Clifford odd objects they resemble the properties of the Clifford even partners 
of the ``basis vectors'', appearing on the left hand side of Eq.~(\ref{allcartaneigenvecAdgen}).

Let us see the application of the operators $S^{0d}$ and $\tilde{S}^{0d}$ on the 
Clifford even ``basis vectors'' on the even dimensional part of the $d=2(2n+1)+1$ space.
The Clifford even ``basis vectors'' must have an even number of nilpotents, which 
means that in $d=2(2n+1)$, we must have at least one projector. To obtain all the 
Clifford even ``basis vectors'' we must apply on these starting Clifford even 
``basis vectors'', presented in Eq.~(\ref{allcartaneigenvecAdgen}) on the left hand 
side, the operators $S^{0d}$ and $\tilde{S}^{0d}$.
\begin{small}
\begin{eqnarray}
\label{allcartaneigenvecAdgen}
 d=&&2(2n+1)+1\, \nonumber\\
 {}^{I}{\bf {\cal A}}^{1 \dagger}_{1} =\stackrel{03}{(+i)}\stackrel{12}{(+)} \stackrel{56}{(+)} \cdots \stackrel{d-2 \, d-1}{[+]} \,,\quad 
&& {}^{I}{\bf {\cal A}}^{1 \dagger}_{2^{{d-1}{2}-1}+1}=\stackrel{03}{[-i]}\stackrel{12}{(+)} \stackrel{56}{(+)}
\cdots \stackrel{d-2 \, d-1}{[+]} \gamma^{d}\,,\nonumber\\
 {}^{I}{\bf {\cal A}}^{2 \dagger}_{1} = \stackrel{03}{[-i]} \stackrel{12}{[-]} 
\stackrel{56}{(+)} \cdots \stackrel{d-2 \, d-1}{[+]}\,, \quad
&&{}^{I}{\bf {\cal A}}^{2 \dagger}_{2^{{d-1}{2}-1}+1} = \stackrel{03}{(+i)} \stackrel{12}{[-]} \stackrel{56}{(+)} \cdots \stackrel{d-2 \, d-1}{[+]} \gamma^{d}\,,\nonumber\\
\cdots \quad 
&& \cdots\nonumber\\
 {}^{I}{\bf {\cal A}}^{2^{\frac{d-1}{2}-1} \dagger}_{1} = \stackrel{03}{[-i]} 
 \stackrel{12}{[-]} \stackrel{56}{[-]} \dots \stackrel{d-2\,d-1}{[+]}\;, \quad
&& {}^{I}{\bf {\cal A}}^{2^{\frac{d-1}{2}-1} \dagger}_{2^{{d-1}{2}-1}+1} = 
\stackrel{03}{(+i)} \stackrel{12}{[-]} 
\stackrel{56}{[-]} \dots \stackrel{d-2\,d-1}{[+]}\; \gamma^{d}\,, \nonumber\\
\cdots \quad
&& \cdots \,.
\end{eqnarray}
\end{small}

The right hand side of Eq.~(\ref{allcartaneigenvecAdgen}), and for the special cases 
of the Clifford even part of Eqs.~( \ref{allcartaneigenvecbd2+1}, 
\ref{allcartaneigenvecbd4+1}), are the Cliffdord even ``basis vectors'' as there are their 
left handed partners. But they resemble properties of the left handed ``basis vectors'';
presented in Eq.~(\ref{allcartaneigenvecbdgen}), and for the special cases 
of the Clifford odd part of Eqs.~( \ref{allcartaneigenvecbd2+1}, 
\ref{allcartaneigenvecbd4+1}). 
These Clifford even objects can be arranged into $2^{\frac{d-1}{2}-1}$ members in $2^{\frac{d-1}{2}-1}$ families of ``basis vectors'' and into  a separate group 
of their Hermitian conjugated partners. However, they are the Clifford even ``basis vectors''.

Let us point out that the Lorentz transformations in internal spaces of fermion and 
boson fields transform the left hand sides of Eq.~((\ref{allcartaneigenvecbdgen}) and 
of Eq.~((\ref{allcartaneigenvecAdgen}) into the corresponding right hand sides and 
opposite.

If we apply algebraically the Clifford  even ``basis vectors'' appearing on the right hand
side of  Eq.~(\ref{allcartaneigenvecAdgen}) on  the Clifford  odd ``basis vectors'' 
appearing on the  right hand side of  Eq.~(\ref{allcartaneigenvecbdgen}), we end up 
with the Clifford odd  ``basis vector'' appearing on the left hand side of  Eq.~(\ref{allcartaneigenvecbdgen}),  or on one of their Hermitian conjugated partners. 
Or we obtain zero.

If we apply algebraically the Clifford  even ``basis vectors'' appearing on the right 
hand side of  Eq.~(\ref{allcartaneigenvecAdgen}) on  the Clifford  odd ``basis vectors'' 
appearing on the  left hand side of  Eq.~(\ref{allcartaneigenvecbdgen}), we end up 
with the Clifford odd  ``basis vectors'' appearing on the right hand side of  Eq.~(\ref{allcartaneigenvecbdgen}).
 
 In the next section, we discuss concrete cases to make discussions more transparent.
 
 \vspace{3mm}
 
 Let us conclude this section with what we have learned:
 
 \vspace{2mm}

{\bf a.} In $d=2n+1$ dimensional spaces, $n=1,2,\dots$, there are two kinds of 
the Clifford odd ``basis vectors'': \\
 {\bf a.i.} The ``basis vectors'' are  products of 
an odd number of nilpotents and the rest of the projectors. These ``basis vectors'' appear 
in $2^{\frac{d-1}{2}-1}$  families, each family has $2^{\frac{d-1}{2}-1}$ members.
They anti-commute, fulfilling together with their Hermitian conjugated partners
the postulates for the second quantized fermion fields. Their Hermitian conjugated
partners appear in a separate group. \\
{\bf a.ii.} Applying on these Clifford odd ``basis vectors'' the operators $S^{0d}$ 
and $\tilde{S}^{0d}$ the additional two times $2^{\frac{d-1}{2}-1}\times$ 
$2^{\frac{d-1}{2}-1}$ of the Clifford odd ``basis vectors'' follow. These Clifford odd 
``basis vectors'' resemble the properties of the Clifford even ``basis vectors'' from the 
case {\bf b.i.} presented below; They form two orthogonal groups. The members of 
each group have their Hermitian conjugated partners within the same group, or they 
are self-adjoint.\\

{\bf b.} In $d=2n+1$ dimensional spaces, $n=1,2,\dots$, there are two kinds of 
the Clifford even ``basis vectors'':  \\
{\bf b.i.} The ``basis vectors'' are  products of even number of nilpotents and 
the rest of the projectors. These ``basis vectors'' appear in two orthogonal groups with
$2^{\frac{d-1}{2}-1}$$\times 2^{\frac{d-1}{2}-1}$ members. Each group have 
their Hermitian conjugated members within their own group, or they are self-adjoint. 
They commute, fulfilling the postulates for the second quantized boson fields, the 
gauge fields of the corresponding fermion fields of the case {\bf a.i.}.  \\
{\bf b.ii.} Applying on these ``basis vectors'' the operators $S^{0d}$ and 
$\tilde{S}^{0d}$ the additional two times $2^{\frac{d-1}{2}-1}\times$ 
$2^{\frac{d-1}{2}-1}$ Clifford even ``basis vectors'' follow. These Clifford even 
``basis vectors'' resemble the properties of the Clifford odd ``basis vectors'' of the case 
{\bf a.i.}; They form two groups with $2^{\frac{d-1}{2}-1}$ members  in each of the $2^{\frac{d-1}{2}-1}$ families. Their Hermitian conjugated partners appear in a 
separate group. But they commute.\\

{\bf c.i.} When Clifford even ``basis vectors'' of the kind {\bf b.i.} algebraically
apply on the Clifford odd ``basis vectors'' of the kind {\bf a.i.} they transfer to the 
Clifford odd  ``basis vectors'' the integer values of the Cartan subalgebra members
($\pm i$, $\pm 1$ or $0$) or they give zero.\\
{\bf c.ii.} When Clifford even basis vectors'' of the kind {\bf b.ii.}  algebraically
apply on the Clifford odd ``basis vectors'' of the kind {\bf a.ii.} they transfer to the
Clifford odd ``basis vectors'' the integer values of the Cartan subalgebra members,
 ($\pm i$, $\pm 1$ or $0$) or they give zero as in the case {\bf c.i.}.\\
 
{\bf d.i.}  While the Clifford odd 
``basis vectors'' in even dimensional spaces have well-defined handedness, since
the operator of handedness is the Clifford even operator, Eq.~(\ref{Gamma}),
the eigenvectors of the operator of handedness in odd dimensional spaces are 
the superposition of  the ``basis vectors'' of the kind {\bf a.i.} and of the kind {\bf a.ii.}.

\section{``Basis vectors'' in even, $d=2n$  for $n=1,2$, and odd, $d=2n+1$  for $n=1,2$, dimensional spaces}
\label{specialcases}

The internal spaces for fermion and boson fields in even and odd dimensional spaces 
for simple cases are discussed: In Subsect.~\ref{2n} for the choices $d=(1+1)$, 
$d=(3+1)$ and in Subsect.~\ref{ODDdspecial} for $d=(0+1)$,  $d=(2+1)$ and 
$d=(4+1)$. This section is  meant as  an illustration of Sect.~\ref{generalproperties}.

In Refs.~\cite{nBled2022,n2022epjc,nIARD2022,nh2021RPPNP,n2021SQ,2020PartIPartII}
the reader can find the definition of the ``basis vectors'' as the eigenstates of the Cartan
subalgebra of the Lorentz algebra in internal spaces of fermion and boson fields. ``Basis 
vectors'' are  written as superposition of the Clifford odd (for fermions) and the Clifford 
even (for bosons) products of $\gamma^a$'s.   ``Basis vectors'' for fermions have either 
left or right handedness, $\Gamma^{d}$ (the handedness is defined in 
Eq.~(\ref{Gamma})), and appear in families (the family quantum numbers are 
determined by  $\tilde{\gamma}^a$'s, with $\tilde{S}^{ab}= \frac{i}{4} 
\{\tilde{\gamma}^a ,\tilde{\gamma}^b\}_{-}$). The Clifford odd ``basis vectors''  
have their Hermitian conjugated partners  in a separate group. ``Basis vectors'' for 
bosons have no families and have their Hermitian conjugated partners within the same 
group, Sect.~\ref{generalproperties}.
 
The ``basis vectors'' in odd dimensional spaces differ in properties from the ``basis 
vectors'' in even dimensional spaces, as we have concluded in the previous Sect.~\ref{generalproperties}. 

Half of  the Clifford odd ``basis vectors'' have properties as in even dimensional 
spaces~\footnote{
The same choice of the Cartan subalgebra members is made in the case $d=(2n+1)$
and in the case of $d=2n$. The Lorentz transformations in the internal space of
fermion and boson fields transform in Eqs.~(\ref{allcartaneigenvecbdgen},
\ref{allcartaneigenvecAdgen})  the left hand sides into the right hand sides and 
opposite.}.
The remaining half  of the Clifford odd ``basis vectors'' gain properties of the Clifford 
even ``basis vectors''. Half of  the Clifford even ``basis vectors'' have properties 
as in even dimensional spaces.  The remaining half  of  the Clifford even ``basis 
vectors'' gain properties of the Clifford odd ``basis vectors''. 
Since the operator of handedness is is the Clifford odd object (it is the product of 
odd number of $\gamma^a$'s), only the superposition of the Clifford odd and the 
Clifford even ``basis  vectors'' have  a definite handedness~\footnote{Correspondingly the 
eigenvectors of the Cartan subalgebra members have both handednesses, 
$\Gamma^{(2n+1)}=\pm 1$.}.
 
%
\subsection{``Basis vectors'' in even dimensional spaces: $d=(1+1), (3+1)$} 
\label{2n}

To illustrate the differences in properties of the internal spaces of fermion and boson fields 
in even and odd dimensional spaces,  simple cases are discussed.  The definition of nilpotents and 
projectors  and the relations among them can be found in Eq.~(\ref{graphic}) and 
App.~\ref{A}.

\vspace{3mm}

{\bf $d=(1+1)$}

\vspace{2mm}

There are $4 \, (2^{d=2})$ ``eigenvectors" of the Cartan subalgebra  
members, Eq.~(\ref{cartancliffevend}),
$S^{01}$ and ${\bf {\cal S}}^{01}$ of the Lorentz algebra $S^{ab}$ and 
${\bf {\cal S}}^{ab}$ $= S^{01} + \tilde{S}^{01}$ ($S^{ab}= \frac{i}{4} 
\{\gamma^a ,\gamma^b\}_{-}$ $\tilde{S}^{ab}= \frac{i}{4} 
\{\tilde{\gamma}^a ,\tilde{\gamma}^b\}_{-}$), representing one Clifford odd 
``basis vector'' $\hat{b}^{ 1 \dagger}_{1}=$ $\stackrel{01}{(+i)}$ (m=1), 
appearing in one family (f=1) and correspondingly one Hermitian conjugated 
partner $\hat{b}^{ 1}_1=$ $\stackrel{01}{(-i)}$~\footnote{
It is our choice which one, $\stackrel{01}{(+i)}$
or $\stackrel{01}{(-i)}$, we choose as the ``basis vector'' $ \hat{b}^{ 1 \dagger}_1$,  
and which one is its Hermitian conjugated partner. The choice of the ``basis vectors''
determines the vacuum state $|\psi_{oc}>$, Eq.~(\ref{vaccliffodd0}). 
For $ \hat{b}^{ 1 \dagger}_1=$$\stackrel{01}{(+i)}$, the vacuum state is 
$|\psi_{oc}>= \stackrel{01}{[-i]}$ 
(due to the requirement that  $\hat{b}^{ 1 \dagger}_1 |\psi_{oc}>$ is nonzero,
while $\hat{b}^{ 1}_1 |\psi_{oc}>$ is zero), which is the Clifford even object.} 
and two Clifford even ``basis vector'' ${}^{I}{\bf {\cal A}}^{1 \dagger}_{1}=
\stackrel{01}{[+i]}$ and ${}^{II}{\bf {\cal A}}^{1 \dagger}_{1}=
\stackrel{01}{[-i]}$, both self-adjoint.

Correspondingly we have, after using  Eqs.~(\ref{gammatildeantiher}, \ref{signature0}), 
two Clifford odd and two Clifford even eigenvectors of the Cartan subalgebra members
 \begin{eqnarray}
 \label{1+1oddeven}
 && {\rm  \;Clifford \;odd}\nonumber\\
 \hat{b}^{ 1 \dagger}_{1}&=&\stackrel{01}{(+i)}\,, \quad 
 \hat{b}^{ 1 }_{1}=\stackrel{01}{(-i)}\,,\nonumber\\
 &&{\rm \;Clifford \;even} \;\nonumber\\
 {}^{I}{\bf {\cal A}}^{1 \dagger}_{1}&=&\stackrel{01}{[+i]}\,, \quad 
{}^{II}{\bf {\cal A}}^{1 \dagger}_{1}=\stackrel{01}{[-i]}\,.
 \end{eqnarray}
 The two Clifford odd ``basis vectors'' are Hermitian conjugated
 to each other.  The choice is made that  $\hat{b}^{ 1 \dagger}_{1}$ is the ``basis 
 vector'',  the second Clifford odd object  is its Hermitian conjugated partner. Defining 
the handedness as $\Gamma^{(1+1)} =\gamma^0 \gamma^1$, Eq.~(\ref{Gamma}),
it follows, using  Eq.~(\ref{graficcliff}), that $\Gamma^{(1+1)}\,
\hat{b}^{ 1 \dagger}_{1}=  \hat{b}^{ 1 \dagger}_{1}$. $\hat{b}^{ 1 \dagger}_{1}$ 
is the right handed ``basis vector''~\footnote{ 
 We could choose left handed  ``basis vectors'' if choosing 
 $\hat{b}^{ 1 \dagger}_{1}=\stackrel{01}{(-i)}$, but the choice of handedness 
would remain only one.}.

 Each of the two Clifford even ``basis vectors'' is self adjoint 
 ($({}^{I,II}{\bf {\cal A}}^{1 \dagger}_{1})^{\dagger}={}^{I,II}{\bf {\cal A}}^{1 \dagger}_{1}$).\\

Let us notice, taking into account Eqs.~(\ref{graficcliff}, \ref{graficcliff1}), that 
\begin{eqnarray}
\label{1+1oddeven1}
 \{ \hat{b}^{ 1}_{1}(\equiv \stackrel{01}{(-i)})*_{A} \hat{b}^{ 1 \dagger}_{1}
(\equiv \stackrel{01}{(+i)}) \} |\psi_{oc}>&=& {}^{II}{\bf {\cal A}}^{1 \dagger}_{1}(\equiv \stackrel{01}{[-i]}) |\psi_{oc}>= |\psi_{oc}>\,,\nonumber\\
\{ \hat{b}^{ 1\dagger}_{1} (\equiv \stackrel{01}{(+i)})*_{A} \hat{b}^{ 1}_{1}
(\equiv \stackrel{01}{(- i)}) \} |\psi_{oc}>&=& 0\, ,\nonumber\\
{}^{I}{\bf {\cal A}}^{1 \dagger}_{1} (\equiv \stackrel{01}{[+i]})\,*_{A}\,
 \hat{b}^{ 1\dagger}_{1}(\equiv \stackrel{01}{(+i)}) |\psi_{oc}> &=&  
 \hat{b}^{ 1\dagger}_{1} (\equiv \stackrel{01}{(+i)}) |\psi_{oc}>\,,\nonumber\\
  {}^{I}{\bf {\cal A}}^{1 \dagger}_{1} (\equiv \stackrel{01}{[+i]}) \,
 \hat{b}^{ 1}_{1}(\equiv \stackrel{01}{(-i)}) |\psi_{oc}> &=& 0\,, \nonumber\\
 {}^{I}{\bf {\cal A}}^{1 \dagger}_{1}\,*_{A}\,{}^{II}{\bf {\cal A}}^{1 \dagger}_{1}
  &=& 0 ={}^{II}{\bf {\cal A}}^{1 \dagger}_{1}\,*_{A}\,
  {}^{I}{\bf {\cal A}}^{1 \dagger}_{1}\,.
  \end{eqnarray}
 
 The case with $d=(3+1)$ is more informative: \\

\vspace{3mm}

{\bf $d=(3+1)$}

\vspace{2mm}
 

In $d=(3+1) $ there are $16 \, (2^{d=4})$ ``eigenvectors" of the Cartan subalgebra  
members ($S^{03}, S^{12}$) and (${\bf {\cal S}}^{03}, {\bf {\cal S}}^{12}$) of the 
Lorentz algebras $S^{ab}$ and ${\bf {\cal S}}^{ab}$ , Eq.~(\ref{cartancliffevend}).

Half of them are the Clifford odd ``basis vectors'', appearing in two  families 
$2^{\frac{4}{2}-1}$, $f=(1,2)$), each with two ($2^{\frac{4}{2}-1}$, $m=(1,2)$), 
members, $\hat{b}^{ m \dagger}_{f}$,  and  $2^{\frac{4}{2}-1}\times $
$2^{\frac{4}{2}-1} $ Hermitian conjugated partners  $\hat{b}^{ m}_{f}$ appearing 
in a separate group (not reachable by $S^{ab}$). 

There are $2^{\frac{4}{2}-1}\times 2^{\frac{4}{2}-1} $  Clifford even ''basis vectors'',
the members of the group ${}^{I}{\bf {\cal A}}^{m \dagger}_{f}$, which are Hermitian conjugated to each other or are self adjoint, all reachable by ${\bf {\cal S}}^{ab}$ 
from any starting ''basis vector'' ${}^{I}{\bf {\cal A}}^{1\dagger}_{1}$. 
And there is another group of $2^{\frac{4}{2}-1}\times 2^{\frac{4}{2}-1} $ Clifford 
even ''basis vectors'', they are the  members of ${}^{II}{\bf {\cal A}}^{m \dagger}_{f}$, 
again either Hermitian conjugated to each other or are self adjoint. All are reachable 
from the starting vector ${}^{II}{\bf {\cal A}}^{1\dagger}_{1}$ by the application 
of ${\bf {\cal S}}^{ab}$. 

 Choosing the right handed  Clifford odd ``basis vectors'' as
\begin{eqnarray}
\label{3+1oddb}
\begin{array} {ccrr}
f=1&f=2&&\\
\tilde{S}^{03}=\frac{i}{2}, \tilde{S}^{12}=-\frac{1}{2}&
\;\;\tilde{S}^{03}=-\frac{i}{2}, \tilde{S}^{12}=\frac{1}{2}\;\;\; &S^{03}\, &S^{12}\\
\hat{b}^{ 1 \dagger}_{1}=\stackrel{03}{(+i)}\stackrel{12}{[+]}&
\hat{b}^{ 1 \dagger}_{2}=\stackrel{03}{[+i]}\stackrel{12}{(+)}&\frac{i}{2}&
\frac{1}{2}\\
\hat{b}^{ 2 \dagger}_{1}=\stackrel{03}{[-i]}\stackrel{12}{(-)}&
\hat{b}^{ 2 \dagger}_{2}=\stackrel{03}{(-i)}\stackrel{12}{[-]}&-\frac{i}{2}&
-\frac{1}{2}\,,
\end{array}
\end{eqnarray}
we find for their Hermitian conjugated partners 
\begin{eqnarray}
\label{3+1oddHb}
\begin{array} {ccrr}
S^{03}=- \frac{i}{2}, S^{12}=\frac{1}{2}&
\;\;S^{03}=\frac{i}{2}, S^{12}=-\frac{1}{2}\;\;&\tilde{S}^{03} &\tilde{S}^{12}\\
\hat{b}^{ 1 }_{1}=\stackrel{03}{(-i)}\stackrel{12}{[+]}&
\hat{b}^{ 1 }_{2}=\stackrel{03}{[+i]}\stackrel{12}{(-)}&-\frac{i}{2}&
-\frac{1}{2}\\
\hat{b}^{ 2 }_{1}=\stackrel{03}{[-i]}\stackrel{12}{(+)}&
\hat{b}^{ 2 }_{2}=\stackrel{03}{(+i)}\stackrel{12}{[-]}&\frac{i}{2}&
\frac{1}{2}\,.
\end{array}
\end{eqnarray}
The vacuum state on which the Clifford odd ''basis vectors apply is equal to:
$|\psi_{oc}>= \frac{1}{\sqrt{2}} (\stackrel{03}{[-i]}\stackrel{12}{[+]}
  +\stackrel{03}{[+i]}\stackrel{12}{[+]} )$~\footnote{
The case $SO(1,1)$ can be viewed as a subspace of the case $SO(3,1)$,
recognizing the ``basis vectors'' $\stackrel{03}{(+i)}\stackrel{12}{[+]}$
and $\stackrel{03}{(-)}\stackrel{12}{[-]}$ with  $\stackrel{03}{(+i)}$ and 
$\stackrel{03}{(-i)}$, respectively, as carrying two different handedness in
$d=(1+1)$, but each of them carries a different ``charge'' $S^{12}$. In the 
whole internal space, all the Clifford odd ``basis 
vectors'' have only one handedness.}. 
  
Let us recognize that all the Clifford odd ''basis  vectors'' are orthogonal: \\
$\hat{b}^{ m \dagger}_{f} *_{A} \hat{b}^{ m' \dagger}_{f '}=0$.

Let us present  $2^{\frac{4}{2}-1}\times 2^{\frac{4}{2}-1} $  Clifford even ''basis vectors'',
the members of the group ${}^{I}{\bf {\cal A}}^{m \dagger}_{f}$, which are Hermitian conjugated to each other or are self adjoint~\footnote{
Let be repeated that ${\bf {\cal S}}^{ab}=S^{ab} + \tilde{S}^{ab} $~\cite{n2022epjc}.}
\begin{eqnarray}
\label{3+1evenAI}
\begin{array} {crrcrr}
&{\bf {\cal S}}^{03}&{\bf {\cal S}}^{12}&&{\bf {\cal S}}^{03}&{\bf {\cal S}}^{12}\\
{}^{I}{\bf {\cal A}}^{1 \dagger}_{1}= \stackrel{03}{[+i]}\stackrel{12}{[+]}&0&0&\,,
{}^{I}{\bf {\cal A}}^{1 \dagger}_{2}= \stackrel{03}{(+i)}\stackrel{12}{(+)}&i&1\\
{}^{I}{\bf {\cal A}}^{2 \dagger}_{1}= \stackrel{03}{(-i)}\stackrel{12}{(-)}&-i&-1&\,,
{}^{I}{\bf {\cal A}}^{2 \dagger}_{2}= \stackrel{03}{[-i]}\stackrel{12}{[-]}&0&0\,,
\end{array}
\end{eqnarray}
and  $2^{\frac{4}{2}-1}\times 2^{\frac{4}{2}-1} $  Clifford even ''basis vectors'',
the members of the group ${}^{II}{\bf {\cal A}}^{m \dagger}_{f}$, $m=(1,2), f=(1,2)$,
which are again Hermitian conjugated to each other or are self adjoint
\begin{eqnarray}
\label{3+1evenAII}
\begin{array} {crrcrr}
&{\bf {\cal S}}^{03}&{\bf {\cal S}}^{12}&&{\bf {\cal S}}^{03}&{\bf {\cal S}}^{12}\\
{}^{II}{\bf {\cal A}}^{1 \dagger}_{1}= \stackrel{03}{[+i]}\stackrel{12}{[-]}&0&0&\,,
{}^{II}{\bf {\cal A}}^{1 \dagger}_{2}= \stackrel{03}{(+i)}\stackrel{12}{(-)}&i&-1\\
{}^{II}{\bf {\cal A}}^{2 \dagger}_{1}= \stackrel{03}{(-i)}\stackrel{12}{(+)}&-i&1&\,,
{}^{II}{\bf {\cal A}}^{2 \dagger}_{2}= \stackrel{03}{[-i]}\stackrel{12}{[+]}&0&0\,.
\end{array}
\end{eqnarray}
The Clifford even ``basis vectors'' have no families. The two groups which are not reachable by 
${\bf {\cal S}}^{ab}$ are orthogonal.
\begin{eqnarray}
\label{AIAIIorth}
{}^{I}{\bf {\cal A}}^{m \dagger}_{f} *_{A} {}^{II}{\bf {\cal A}}^{m' \dagger}_{f `} 
=0, \quad{\rm for \;any } \;(m, m', f, f `)\,.
\end{eqnarray}
Even dimensional spaces have the properties of the fermion and boson second quantized 
fields. The reader can find discussions about $d=(5+1)$- dimensional case in~\cite{n2022epjc,nh2021RPPNP} and the references therein.


%
\subsection{``Basis vectors''  in odd dimensional spaces with $d=(2+1),
(4+1)$}
\label{ODDdspecial}

\vspace{3mm}

Half of the Clifford odd and Clifford even Clifford objects in $2n+1$-%
dimensional cases can be found by treating  the Clifford odd ``basis 
vectors'' and their Hermitian conjugated partners and the Clifford even ``basis 
vectors'' in $2(2n+1)$ (or $4n$) dimensional part of space. The properties of
these ``basis vectors'' are presented in Eqs.~(\ref{allcartaneigenvecb}, 
\ref{allcartaneigenvecb4n}, \ref{allcartaneigenvecevenI}, \ref{allcartaneigenvecevenII}). 

The rest of  the ``basis vectors'' follow by the application of $S^{0 d}$  on the ``basis 
vectors'' determining the internal space of fermion and boson fields in $2(2n+1)$ 
(or $4n$) dimensional part of space. Since $S^{0 d}$ are the Clifford even operators,   
they do not change oddness or evenness of the  ``basis vectors'' or their Hermitian 
conjugated partners. But they do change their properties:\\
{\bf i. } In even dimensional subspace, $2(2n+1)$  of $d=2(2n+1)+1)$ (or $4n$ of 
$d=4n+1$) the Clifford odd ``basis vectors'', $\hat{b}^{m \dagger}_{f}$,  have 
$2^{\frac{d-1}{2}-1}$ members, $m$, in $2^{\frac{d-1}{2}-1}$ families, $f$, and their Hermitian conjugated partners appear in a separate group of $2^{\frac{d-1}{2}-1}$ 
members in $2^{\frac{d-1}{2}-1}$ families.  The Clifford even ``basis vectors'' appear
in two mutually orthogonal groups, each with $2^{\frac{d-1}{2}-1}\times$
$2^{\frac{d-1}{2}-1}$ members. \\ 
{\bf ii. } The second part of ``basis vectors'' and their Hermitian conjugated partners, 
obtained from the first part by the application of $S^{0 d}$ with the same number of 
either the Clifford odd or of the Clifford even objects as the first part, manifest:\\ 
The Clifford odd ``basis vectors'' appear in two mutually orthogonal groups, each with $2^{\frac{d-1}{2}-1}\times$ $2^{\frac{d-1}{2}-1}$ members, self adjoint or with the 
Hermitian conjugated partners within the same group.  The Clifford even
``basis vectors'' appear in $2^{\frac{d-1}{2}-1}$ members, $m$, in $2^{\frac{d-1}{2}-1}$ families, $f$, and their Hermitian conjugated partners appear in a separate group of $2^{\frac{d-1}{2}-1}$ members in $2^{\frac{d-1}{2}-1}$ families. \\
{\bf iii.}  While $\hat{b}^{m \dagger}_{f}$ have in even dimensional spaces one 
handedness only (either right or left, depending on the definition of handedness), 
in odd dimensional spaces, the operator of handedness is a Clifford odd object --- 
the product of an odd number of $\gamma^a$'s, Eq.~(\ref{Gamma}), (still 
commuting with $S^{ab}$) --- transforming the Clifford odd ``basis vectors'' into 
Clifford even ``basis vectors'' and opposite. 
Correspondingly are the eigenvectors of the operator of handedness the 
superposition of the Clifford odd and the Clifford even basis vectors'', offering in 
odd dimensional spaces the right and left handed eigenvectors of the operator 
of handedness.  \\

Let us illustrate the above mentioned properties of the ``basis vectors'' in odd 
dimensional spaces, starting with the simplest case: 

\vspace{3mm}

{\bf d=(2+1)}

\vspace{2mm}

In $d=(2+1) $ there are $8 \, (2^{d=3})$ ``eigenvectors" of the Cartan subalgebra  
members ($S^{01}$) and (${\bf {\cal S}}^{01}$) of the 
Lorentz algebras $S^{ab}$ and ${\bf {\cal S}}^{ab}$ , Eq.~(\ref{cartancliffODDd}).

Half of the Clifford odd  and Clifford even ``basis vectors'' and their Hermitian conjugated
partners can be taken from Eq.~(\ref{1+1oddeven}), the rest half are obtained by the 
application of $S^{02}$ or $\tilde{S}^{02}$ on the first half. One obtains
%
\begin{small}
\begin{eqnarray}
\label{allcartaneigenvecbd2+1}
 d=&&2+1\, \nonumber\\
 && {\rm Clifford\;  odd} \nonumber\\
 \hat{b}^{1 \dagger}_{1}=\stackrel{01}{(+i)}\,,\quad 
&& \hat{b}^{1 \dagger}_{2}=\stackrel{01}{[-i]}\gamma^{2}\,,\nonumber\\
\hat{b}^{1 }_{1} = \stackrel{01}{(-i)}\,, \quad
&&\hat{b}^{1}_{2} = \stackrel{01}{[+i]}  \gamma^{2}\,,\nonumber\\
&& \nonumber\\
&& {\rm Clifford\;  even} \nonumber\\
{}^{I}{\bf {\cal A}}^{1\dagger}_{1}= \stackrel{01}{[+i]} \,,\quad
&&{}^{I}{\bf {\cal A}}^{1\dagger}_{2}= \stackrel{01}{(-i)} \gamma^2\,, 
\nonumber\\
{}^{II}{\bf {\cal A}}^{1\dagger}_{1}= \stackrel{01}{[-i]} \,,\quad
&&{}^{II}{\bf {\cal A}}^{1\dagger}_{2}= \stackrel{01}{(+i)} \gamma^2\,.
\end{eqnarray}
\end{small}
One clearly sees that the left hand side of the Clifford odd ``basiss vectors'' and the 
right hand side of the Clifford even ``basis vectors'', although the  first are the Clifford 
odd objects and the later Clifford even objects, have similar properties. 

Like:
$$\hat{b}^{1 }_{1} *_A  \hat{b}^{1 \dagger}_{1}= 
{}^{I}{\bf {\cal A}}^{1\dagger}_{2} 
*_A  {}^{II}{\bf {\cal A}}^{1\dagger}_{2}= \stackrel{01}{(-i)} \stackrel{01}{(+i)} =
\stackrel{01}{[-i]} =|\psi_{oc}>\,.$$

And the right hand side of the Clifford odd ``basis vectors'' contains two self adjoint 
orthogonal ``basis vectors'' as the left hand side of the two Clifford even ``basis
vectors'' does.

Let us find the eigenvectors of the operator of handedness 
$\Gamma^{(2+1)}=i\gamma^0\gamma^1\gamma^2$. Since it is the Clifford odd
object, its eigenvectors are the superposition of Clifford odd and Clifford even ``basis vectors''.
\begin{small}
\begin{eqnarray}
\label{Gamma21}
\begin{array} {c}
\Gamma^{(2+1)}\{\stackrel{01}{[-i]}\pm i \stackrel{01}{[-i]}\gamma^2\} =
\mp \{\stackrel{01}{[-i]}\pm i \stackrel{01}{[-i]}\gamma^2\}\,,\\
\Gamma^{(2+1)}\{\stackrel{01}{(+i)}\pm i \stackrel{01}{(+i)}\gamma^2\} =
\mp\{\stackrel{01}{(+i)}\pm i \stackrel{01}{(+i)}\gamma^2\}\,,\\
\Gamma^{(2+1)}\{\stackrel{01}{[+i]}\pm i \stackrel{01}{[+i]}\gamma^2\} =
\pm \{\stackrel{01}{[+i]}\pm i \stackrel{01}{[+i]}\gamma^2\}\,,\\
\Gamma^{(2+1)}\{\stackrel{01}{(-i)}\gamma^2 \pm i \stackrel{01}{(-i)}\} =
\pm \{\stackrel{01}{(-i)}\gamma^2 \pm i \stackrel{01}{(-i)}\}\,.\\
\end{array}
\end{eqnarray}
\end{small}

\vspace{3mm}

{\bf d=(4+1)}

\vspace{2mm}

In $d=(4+1) $ there are $32 \, (2^{d=5})$ ``eigenvectors" of the Cartan subalgebra  
members ($S^{03}, S^{12}$) and (${\bf {\cal S}}^{03}, {\bf {\cal S}}^{12}$) of the 
Lorentz algebras $S^{ab}$ and ${\bf {\cal S}}^{ab}$, Eq.~(\ref{cartancliffODDd}).

Half of the Clifford odd  and Clifford even ``basis vectors'' and their Hermitian conjugated
partners can be taken from Eqs.~(\ref{3+1oddb}, \ref{3+1oddHb}, \ref{3+1evenAI},
\ref{3+1evenAII}),  the rest half follows by the application of $S^{05}$ or 
$\tilde{S}^{05}$ on the first half. 

\begin{small}
\begin{eqnarray}
\label{allcartaneigenvecbd4+1}
 d=&&4+1\, \nonumber\\
 && {\rm Clifford\;  odd} \nonumber\\
 \hat{b}^{1 \dagger}_{1}=\stackrel{03}{(+i)}\stackrel{12}{[+]}\,,\;\, 
 \hat{b}^{1 \dagger}_{2}=\stackrel{03}{[+i]}\stackrel{12}{(+)}\,,
 &&\hat{b}^{1\dagger}_{3} =\stackrel{03}{[-i]}\stackrel{12}{[+i]
 } \gamma^{5}\,,
 \; \, \hat{b}^{1 \dagger}_{4} =\stackrel{03}{(-i)}\stackrel{12}{(+)}\gamma^5 \,,\nonumber\\
\hat{b}^{2 \dagger}_{1}=\stackrel{03}{[-i]}\stackrel{12}{(-)}\,,\;\, 
 \hat{b}^{2 \dagger}_{2}=\stackrel{03}{(-i)}\stackrel{12}{[-]}\,,
 &&\hat{b}^{2\dagger}_{3}=\stackrel{03}{(+i)}\stackrel{12}{(-)} \gamma^{5}\,, \; \, 
\hat{b}^{2\dagger}_{4}= \stackrel{03}{[+i]}\stackrel{12}{[-]}\gamma^5 \,,\nonumber\\
\hat{b}^{1}_{1} = \stackrel{03}{(-i)}\stackrel{12}{[+]} \,, \;\,
\hat{b}^{1}_{2} = \stackrel{03}{[+i]} \stackrel{12}{(-)} \,,
&&\hat{b}^{1}_{3} = \stackrel{03}{[+i]}\stackrel{12}{[+]} \gamma^{5}\,,\;\,
\hat{b}^{1}_{4}= \stackrel{03}{(-i)}\stackrel{12}{(-)} \gamma^5 \,,\nonumber\\
\hat{b}^{2}_{1} = \stackrel{03}{[-i]}\stackrel{12}{(+)} \,, \;\,
\hat{b}^{2}_{2} = \stackrel{03}{(+i)} \stackrel{12}{[-]} \,,
&&\hat{b}^{2}_{3} = \stackrel{03}{(+i)}\stackrel{12}{(+)} \gamma^{5}\,,\;\,
\hat{b}^{2}_{4}= \stackrel{03}{[-i]}\stackrel{12}{[-]} \gamma^5\,,\nonumber\\
&& \nonumber\\
&& {\rm Clifford\;  even} \nonumber\\
{}^{I}{\bf {\cal A}}^{1\dagger}_{1}= \stackrel{03}{[+i]}\stackrel{12}{[+]} \,,\;\,
{}^{I}{\bf {\cal A}}^{1\dagger}_{2}= \stackrel{03}{(+i)}\stackrel{12}{(+)} \,,
&&{}^{I}{\bf {\cal A}}^{1}_{3}= \stackrel{03}{(-i)}\stackrel{12}{[+]}\gamma^5\,,\;\,
{}^{I}{\bf {\cal A}}^{1}_{4}= \stackrel{03}{[-i]}\stackrel{12}{(+)}\gamma^5\,, 
\nonumber\\
{}^{I}{\bf {\cal A}}^{2\dagger}_{1}= \stackrel{03}{(-i)}\stackrel{12}{(-i)} \,,\;\,
{}^{I}{\bf {\cal A}}^{2\dagger}_{2}= \stackrel{03}{[-i]}\stackrel{12}{[-]} \,,
&&{}^{I}{\bf {\cal A}}^{2}_{3}= \stackrel{03}{[+i]}\stackrel{12}{(-)}
\gamma^5\,,\;\,
{}^{I}{\bf {\cal A}}^{2}_{4}= \stackrel{03}{(+i)}\stackrel{12}{[-]}\, \gamma^5\,, 
\nonumber\\
&&\nonumber\\
{}^{II}{\bf {\cal A}}^{1\dagger}_{1}= \stackrel{03}{[-i]}\stackrel{12}{[+]} \,,\;\,
{}^{II}{\bf {\cal A}}^{1\dagger}_{2}= \stackrel{03}{(-i)}\stackrel{12}{(+)} \,,
&&{}^{II}{\bf {\cal A}}^{1\dagger}_{3}= \stackrel{03}{(+i)}\stackrel{12}{[+]}
\gamma^5\,,\;\,
{}^{II}{\bf {\cal A}}^{1\dagger}_{4}= \stackrel{03}{[+i]}\stackrel{12}{(+)}\gamma^5\,, 
\nonumber\\
{}^{II}{\bf {\cal A}}^{2\dagger}_{1}= \stackrel{03}{(+i)}\stackrel{12}{(-)} \,,\;\,
{}^{II}{\bf {\cal A}}^{2\dagger}_{2}= \stackrel{03}{[+i]}\stackrel{12}{[-]} \,,
&&{}^{II}{\bf {\cal A}}^{2\dagger}_{3}= \stackrel{03}{[-i]}\stackrel{12}{(-)}
\gamma^5\,,\;\,
{}^{II}{\bf {\cal A}}^{2\dagger}_{4}= \stackrel{03}{(-i)}\stackrel{12}{[-]}\,
\gamma^5 \,.
\end{eqnarray}
\end{small}
%

One notices that the right hand side of the Clifford odd ``basis vectors'' appear in two 
mutually orthogonal groups,  each one with either self-adjoint members or with the
Hermitian conjugated partners within the same group. 

The members of one group  
$$\hat{b}^{1\dagger}_{3} =\stackrel{03}{[-i]}\stackrel{12}{[+i] } \gamma^{5}\,,\;
 \hat{b}^{1 \dagger}_{4} =\stackrel{03}{(-i)}\stackrel{12}{(+)}\gamma^5\,, \; 
\hat{b}^{2\dagger}_{3}=\stackrel{03}{(+i)}\stackrel{12}{(-)} \gamma^{5}\,,  \;
\hat{b}^{2\dagger}_{4}= \stackrel{03}{[+i]}\stackrel{12}{[-]}\gamma^5 $$
have the properties, except the commutativity (they are namely, the Clifford odd
objects), as the group of  Clifford even objects
$${}^{II}{\bf {\cal A}}^{1\dagger}_{1}= \stackrel{03}{[-i]}\stackrel{12}{[+]} \,,\;
{}^{II}{\bf {\cal A}}^{1\dagger}_{2}= \stackrel{03}{(-i)}\stackrel{12}{(+)}\,,\;
{}^{II}{\bf {\cal A}}^{2\dagger}_{1}= \stackrel{03}{(+i)}\stackrel{12}{(-)} \,,\;
{}^{II}{\bf {\cal A}}^{2\dagger}_{2}= \stackrel{03}{[+i]}\stackrel{12}{[-]} \,.$$
The comparable properties also have the Clifford odd members of the group
$$\hat{b}^{1}_{3} = \stackrel{03}{[+i]}\stackrel{12}{[+]} \gamma^{5}\,,\;
\hat{b}^{1}_{4}= \stackrel{03}{(-i)}\stackrel{12}{(-)} \gamma^5\,,\;
\hat{b}^{2}_{3} = \stackrel{03}{(+i)}\stackrel{12}{(+)} \gamma^{5}\,,\;
\hat{b}^{2}_{4}= \stackrel{03}{[-i]}\stackrel{12}{[-]} \gamma^5\,,$$
and the Clifford even members of the group
$${}^{I}{\bf {\cal A}}^{1\dagger}_{1}= \stackrel{03}{[+i]}\stackrel{12}{[+]} \,,\;
{}^{I}{\bf {\cal A}}^{1\dagger}_{2}= \stackrel{03}{(+i)}\stackrel{12}{(+)} \,,\;
{}^{I}{\bf {\cal A}}^{2\dagger}_{1}= \stackrel{03}{(-i)}\stackrel{12}{(-i)} \,,\;
{}^{I}{\bf {\cal A}}^{2\dagger}_{2}= \stackrel{03}{[-i]}\stackrel{12}{[-]} \,.$$
The members of both groups have Hermitian conjugated partners within the same 
group or are self-adjoint.

On the other side, the members of the Clifford even group
$${}^{II}{\bf {\cal A}}^{1\dagger}_{3}= \stackrel{03}{(+i)}\stackrel{12}{[+]}
\gamma^5\,,\;
{}^{II}{\bf {\cal A}}^{1\dagger}_{4}= \stackrel{03}{[+i]}\stackrel{12}{(+)}\gamma^5\,,\;
{}^{II}{\bf {\cal A}}^{2\dagger}_{3}= \stackrel{03}{[-i]}\stackrel{12}{(-)}
\gamma^5\,,\;
{}^{II}{\bf {\cal A}}^{2\dagger}_{4}= \stackrel{03}{(-i)}\stackrel{12}{[-]}\, 
\gamma^5\,,$$
have their Hermitian conjugated partners in a separate group 
$${}^{I}{\bf {\cal A}}^{1}_{3}= \stackrel{03}{(-i)}\stackrel{12}{[+]}\gamma^5\,\;\,
{}^{I}{\bf {\cal A}}^{1}_{4}= \stackrel{03}{[+i]}\stackrel{12}{(-)}\gamma^5\,, 
{}^{I}{\bf {\cal A}}^{2}_{3}= \stackrel{03}{[-i]}\stackrel{12}{(+)}\gamma^5\,,
{}^{I}{\bf {\cal A}}^{2}_{4}= \stackrel{03}{(+i)}\stackrel{12}{[-]}\, \gamma^5\,,$$
just like the Clifford odd objects on the left hand side 
$$ \hat{b}^{1 \dagger}_{1}=\stackrel{03}{(+i)}\stackrel{12}{[+]}\,, \;
 \hat{b}^{1 \dagger}_{2}=\stackrel{03}{[+i]}\stackrel{12}{(+)}\,,\;
\hat{b}^{2 \dagger}_{1}=\stackrel{03}{[-i]}\stackrel{12}{(-)}\,, \;
 \hat{b}^{2 \dagger}_{2}=\stackrel{03}{(-i)}\stackrel{12}{[-]}\,,$$
which have their Hermitian conjugated partners in a separate group
$$\hat{b}^{1}_{1} = \stackrel{03}{(-i)}\stackrel{12}{[+]} \,, \;
\hat{b}^{1}_{2} = \stackrel{03}{[+i]} \stackrel{12}{(-)} \,,\;
\hat{b}^{2}_{1} = \stackrel{03}{[-i]}\stackrel{12}{(+)} \,, \;
\hat{b}^{2}_{2} = \stackrel{03}{(+i)} \stackrel{12}{[-]} \,.$$

The ``basis vectors'' of the right hand side  keep oddness if they are partners of 
the Clifford odd ``basis vectors'' on left hand side,  but demonstrate properties of 
the Clifford even objects on the left hand side.
 
The ``basis vectors'' of the right hand side  keep evenness if they are partners of 
the Clifford even ``basis vectors'' on the left hand side,  but demonstrate properties 
of the Clifford odd objects on the left hand side.

After algebraically application of, for example,  ${}^{II}{\bf {\cal A}}^{1\dagger}_{3}
(= \stackrel{03}{(+i)}\stackrel{12}{[+]} \gamma^5 $ on 
$\hat{b}^{1 \dagger}_{4} =\stackrel{03}{(-i)}\stackrel{12}{(+)}\gamma^5 $ 
we are left with $\hat{b}^{1 \dagger}_{2} =\stackrel{03}{[+i]}\stackrel{12}{(+)}$.

The eigenvectors of the operator of handedness in $d=(4+1)$, 
$\Gamma^{(4+1)}=$ $\gamma^0 \gamma^1 \gamma^2\gamma^3 \gamma^5 $,
are the superposition of the Clifford odd and Clifford even  ``basis vectors'', as for example:
$\Gamma^{(4+1)} (\hat{b}^{1 \dagger}_{1} [= \stackrel{03}{(+i)}
\stackrel{12}{[+]}] \pm {}^{II}{\bf {\cal A}}^{1\dagger}_{3}
[= \stackrel{03}{(+i)}\stackrel{12}{[+]} \gamma^5 ] )=
\mp ((\hat{b}^{1 \dagger}_{1} \pm  {}^{II}{\bf {\cal A}}^{1\dagger}_{3})$.

\vspace{2mm}

We can conclude that neither Clifford odd nor Clifford even ``basis vectors'', 
have in odd dimensional spaces the properties which they do demonstrate 
in even dimensional spaces: The properties which empower the Clifford odd 
``basis vectors'' to describe the internal space of fermion fields and the 
Clifford even ``basis vectors'' to describe the internal space of the 
corresponding gauge fields:  After enlarging the ``basis vectors'' in a tensor 
product, $*_{T}$,  with the basis in ordinary space~\cite{n2022epjc}, the
corresponding creation and annihilation operators manifest the properties
required by the postulates for the  second quantized either fermion or boson 
fields, respectively.

In odd dimensional spaces, half of the Clifford odd ``basis vectors'' demonstrate
properties of the Clifford even ``basis vectors'' and half of the Clifford  even 
``basis vectors'' demonstrate properties of the Clifford odd ``basis vectors''.
Arbitrary Lorentz transformations transform  the left hand sides into the right 
sides and vice versa.

These are properties of the internal spaces of the ghost scalar fields, used
in the quantum field theory to make contributions of the Feynman diagrams 
finite.\\

\section{Discussion}

\label{discussion}

This article discusses the properties of the internal spaces  of fermion and boson 
fields in even and odd dimensional spaces, if the internal spaces are described
by the Clifford odd and even ``basis vectors'', which are the superposition of odd 
or even products of operators $\gamma^a$'s. ``Basis vectors'' are arranged into
algebraic products of nilpotents and projectors, which are eigenvectors of the
Cartan subalgebra of the Lorentz algebra $S^{ab}$ in the internal space of 
fermion and bosons fields.

The Clifford odd ``basis vectors'', which are products of an odd number of nilpotents 
and the rest of projectors, offer in even dimensional spaces the description of the 
internal space of fermion fields.\\
Each irreducible representation of the Lorentz algebra is equipped with the family
quantum number determined by the second kind of the Clifford operators 
$\tilde{\gamma}^a$'s. The Clifford odd ``basis vectors'' anti-commute. Their 
Hermitian conjugated partners appear in a different group. In a tensor product
with the basis in ordinary space, the ``basis vectors'' and their Hermitian conjugated
partners form the creation and annihilation operators which, applied on the 
vacuum state or on the Hilbert space~(\cite{nh2021RPPNP} and the references 
therein),  fulfil the anti-commutation relations postulated for the second quantized 
fermion fields, offering therefore the explanation for the postulates. \\
In $d=2(2n+1), n\ge 7$, the creation and annihilation operators, applying on 
the vacuum state, or  the Hilbert space, offer the description of all the properties 
of the observed quarks and leptons and antiquarks and 
antileptons~(\cite{nh2021RPPNP} and the references therein)~\footnote{
Quarks and leptons and antiquarks and antileptons appear in the same irreducible
representation}. 
\vspace{2mm} 

The Clifford even ``basis vectors'', which are products of an even number of nilpotents 
and the rest of projectors offer in even dimensional spaces the description of the 
internal space of boson fields, the gauge fields of the corresponding fermion fields, 
described by the Clifford odd ``basis vectors''.
The Clifford even ``basis vectors'' commute. They do not appear in families and have 
their Hermitian conjugated partners in the same group or are self-adjoint. In a tensor 
product with the basis in ordinary space, the Clifford even ``basis vectors'' form the 
creation and annihilation operators, which fulfil the commutation relations postulated 
for the second quantized boson fields. In $d=2(2n+1), n\ge 7$,
these creation and annihilation operators offer the description of all the properties 
of the observed gauge fields as well as of  Higgs's  scalar field, explaining also the
Yukawa couplings.

This way of describing the internal space of boson fields with the Clifford even ``basis
vectors'', although very promising, needs further studies to understand what new it
can bring into understanding of the second quantization of fermion and boson fields. 
In particular, it must be understood what new, if anything, does  bring the replacement 
in a simple starting action  in $d=2(2n+1), n\ge 7$, Eq.~(\ref{wholeaction}), of vielbeins,
$f^a{}_{\alpha}$, and the two kinds of the spin connection fields, 
$\omega_{ab \alpha}$ (the gauge fields of $S^{ab}$) and $\tilde{\omega}_{ab \alpha}$  
(the gauge fields of $\tilde{S}^{ab}$) 
in the covariant derivative 
$ p_{0 \alpha }$
  $$p_{0\alpha} = p_{\alpha}  - \frac{1}{2}  S^{ab} \omega_{ab \alpha} - 
                    \frac{1}{2}  \tilde{S}^{ab}   \tilde{\omega}_{ab \alpha} \,,
                    \quad \quad \quad\quad\;$$
with 
$$ p_{0\alpha}  = p_{\alpha}  - 
\sum_{m f}   {}^{I}{ \hat {\cal A}}^{m \dagger}_{f}
{}^{I}{\cal C}^{m}_{f \alpha}   - 
 \sum_{m f} {}^{II}{\hat {\cal A}}^{m \dagger}_{f}\,
{}^{I}{\cal C}^{m}_{f \alpha}\,. $$

%
The relations among ${}^{I}{\hat {\cal A}}^{m \dagger}_{f}
{}^{I}{\cal C}^{m}_{f \alpha}$ and $\omega_{ab \alpha}$, and
${}^{II}{\hat {\cal A}}^{m \dagger}_{f}\,
{}^{II}{\cal C}^{m}_{f \alpha}$ 
 and $\tilde{\omega}_{ab \alpha}$, not discussed directly in this article~\cite{n2022epjc}, 
 need additional study.

Not only that the description of the internal spaces of the fermion and boson fields with
the Clifford odd and Clifford even ``basis vectors'' in even dimensional spaces offers an
explanation for the second quantized postulates for fermion and boson fields,  for 
all the assumptions of the {\it standard model}, and for several so far observed 
phenomena,  making several predictions, also the description of the internal spaces 
of the fermion and boson fields in odd dimensional spaces seems meaningful for an
explanation for the ghosts, postulated by Fadeev and Popov~\cite{FadeevPopov}. 
introduced into gauge quantum field theories to take care of the consistency of the 
path integral formulation of the quantum field theory. 

\vspace{2mm}

 Let us repeat what we have learned in this paper, Subsect.~\ref{ODDd},  Subsect.~\ref{ODDdspecial}, about  
 properties of the Clifford even  and the Clifford odd objects  in odd dimensional spaces:\\
Neither Clifford odd nor Clifford even ``basis vectors'' have in odd dimensional spaces 
the properties which they do demonstrate in even dimensional spaces, the properties 
which empower the Clifford odd ``basis vectors'' to describe the internal space of 
fermion fields and the Clifford even ``basis vectors'' to describe the internal space of 
the corresponding gauge fields.

In odd dimensional spaces, namely, half of the Clifford odd ''basis vectors'', 
although anticommuting, demonstrate properties of the Clifford even 
``basis vectors'' in even dimensional spaces and half of the Clifford 
even ``basis vectors'', although commuting, demonstrate properties of the Clifford 
odd ``basis vectors'' in even dimensional spaces.
These ``basis vectors'' obviously resemble properties of the internal spaces of the 
ghost scalar fields, used in the quantum field theory to make contributions of the 
Feynman diagrams finite~\footnote{Arbitrary Lorentz transformations 
in odd dimensional spaces transform  the left hand sides of 
Eqs.~(\ref{allcartaneigenvecbdgen}, \ref{allcartaneigenvecAdgen},
\ref{allcartaneigenvecbd2+1}, \ref{allcartaneigenvecbd4+1}) into the right 
sides and vice versa.}.
These are properties of the internal spaces of the ghost scalar fields used
in the quantum field theory to make contributions of the Feynman diagrams 
finite.

Also, properties of the Clifford odd and the Clifford even ''basis vectors'' in odd 
dimensional spaces need further study.

\appendix

\section{Some useful formulas}
\label{A}

This appendix contains helpful relations needed in this paper. For more detailed 
explanations, and for proofs, the reader is kindly asked to read~\cite{nh2021RPPNP} 
and the references therein.

\vspace{2mm} 
 
 The operator of handedness $\Gamma^d$ is for fermions determined as follows. 
\begin{eqnarray}
\label{Gamma}
 \Gamma^{(d)}= \prod_a (\sqrt{\eta^{aa}} \gamma^a)  \cdot \left \{ \begin{array}{l l}
 (i)^{\frac{d}{2}} \,, &\rm{ for\, d \,even}\,,\\
 (i)^{\frac{d-1}{2}}\,,&\rm{for \, d \,odd}\,,
  \end{array} \right.
 \end{eqnarray}
%

The Clifford objects  $\gamma^a$'s and $\tilde{\gamma}^a$'s fulfil
the relations
\begin{eqnarray}
\label{gammatildeantiherA}
\{\gamma^{a}, \gamma^{b}\}_{+}&=&2 \eta^{a b}= \{\tilde{\gamma}^{a}, 
\tilde{\gamma}^{b}\}_{+}\,, \nonumber\\
\{\gamma^{a}, \tilde{\gamma}^{b}\}_{+}&=&0\,,\quad
 (a,b)=(0,1,2,3,5,\cdots,d)\,, \nonumber\\
(\gamma^{a})^{\dagger} &=& \eta^{aa}\, \gamma^{a}\, , \quad 
(\tilde{\gamma}^{a})^{\dagger} =  \eta^{a a}\, \tilde{\gamma}^{a}\,.
\end{eqnarray}
In the paper the signature $\eta^{aa}=diag(1,-1,-1,\dots,-1)$ is used.

\vspace{3mm}

The choice of the Cartan subalgebra members is made for $d$ even
\begin{eqnarray}
&&{\cal {\bf S}}^{03}, {\cal {\bf S}}^{12}, {\cal {\bf S}}^{56}, \cdots, 
{\cal {\bf S}}^{d-1 \;d}\,, \nonumber\\
&&S^{03}, S^{12}, S^{56}, \cdots, S^{d-1 \;d}\,, \nonumber\\
&&\tilde{S}^{03}, \tilde{S}^{12}, \tilde{S}^{56}, \cdots,  \tilde{S}^{d-1\; d}\,, 
\nonumber\\
&&{\cal {\bf S}}^{ab} = S^{ab} +\tilde{S}^{ab}\,,
\label{cartancliffevend}
\end{eqnarray}
and for $d$ odd
\begin{eqnarray}
&&{\cal {\bf S}}^{03}, {\cal {\bf S}}^{12}, {\cal {\bf S}}^{56}, \cdots, 
{\cal {\bf S}}^{d-2 \;d-1}\,, \nonumber\\
&&S^{03}, S^{12}, S^{56}, \cdots, S^{d-2 \;d-1}\,, \nonumber\\
&&\tilde{S}^{03}, \tilde{S}^{12}, \tilde{S}^{56}, \cdots,  \tilde{S}^{d-2\; d-1}\,, 
\nonumber\\
&&{\cal {\bf S}}^{ab} = S^{ab} +\tilde{S}^{ab}\,.
\label{cartancliffODDd}
\end{eqnarray}

Nilpotents and projectors are defined as follows~\cite{norma93,nh02,nh03}
\begin{eqnarray}
\label{graficcliff}
\stackrel{ab}{(k)}:&=& 
\frac{1}{2}(\gamma^a + \frac{\eta^{aa}}{ik} \gamma^b)\,,\quad 
\stackrel{ab}{[k]}:=\frac{1}{2}(1+ \frac{i}{k} \gamma^a \gamma^b)\,,
\end{eqnarray}
with  $k^2=\eta^{aa} \eta^{bb}$.\\
One finds, taking Eq.~(\ref{gammatildeantiher}) into account, and assuming
\begin{eqnarray}
\{\tilde{\gamma}^a B &=&(-)^B\, i \, B \gamma^a\}\, |\psi_{oc}>\,,
\label{tildegammareduced}
\end{eqnarray}
with $(-)^B = -1$, if $B$ is (a function of) an odd products of $\gamma^a$'s,  otherwise 
$(-)^B = 1$~\cite{nh03}, $|\psi_{oc}>$ is defined in Eq.~(\ref{vaccliffodd}), 
the eigenvalues of the Cartan subalgebra operators 
\begin{eqnarray}
\label{signature0}
S^{ab} \,\stackrel{ab}{(k)} = \frac{k}{2}  \,\stackrel{ab}{(k)}\,,\quad && \quad
\tilde{S}^{ab}\,\stackrel{ab}{(k)} = \frac{k}{2}  \,\stackrel{ab}{(k)}\,,\nonumber\\
S^{ab}\,\stackrel{ab}{[k]} =  \frac{k}{2}  \,\stackrel{ab}{[k]}\,,\quad && \quad 
\tilde{S}^{ab} \,\stackrel{ab}{[k]} = - \frac{k}{2}  \,\,\stackrel{ab}{[k]}\,.
\end{eqnarray}
The vacuum state for the Clifford odd ''basis vectors'', $|\psi_{oc}>$, is defined as
\begin{eqnarray}
\label{vaccliffodd}
|\psi_{oc}>= \sum_{f=1}^{2^{\frac{d}{2}-1}}\,\hat{b}^{m}_{f}{}_{*_A}
\hat{b}^{m \dagger}_{f} \,|\,1\,>\,.
\end{eqnarray}

\vspace{2mm}

Taking into account Eq.~(\ref{gammatildeantiher}) it follows
\begin{small}
\begin{eqnarray}
%
\gamma^a \stackrel{ab}{(k)}&=& \eta^{aa}\stackrel{ab}{[-k]},\; \quad
\gamma^b \stackrel{ab}{(k)}= -ik \stackrel{ab}{[-k]}, \; \quad 
\gamma^a \stackrel{ab}{[k]}= \stackrel{ab}{(-k)},\;\quad \;\;
\gamma^b \stackrel{ab}{[k]}= -ik \eta^{aa} \stackrel{ab}{(-k)}\,,\nonumber\\
\tilde{\gamma^a} \stackrel{ab}{(k)} &=& - i\eta^{aa}\stackrel{ab}{[k]},\quad
\tilde{\gamma^b} \stackrel{ab}{(k)} =  - k \stackrel{ab}{[k]}, \;\qquad  \,
\tilde{\gamma^a} \stackrel{ab}{[k]} =  \;\;i\stackrel{ab}{(k)},\; \quad
\tilde{\gamma^b} \stackrel{ab}{[k]} =  -k \eta^{aa} \stackrel{ab}{(k)}\,, 
\nonumber\\
\stackrel{ab}{(k)}^{\dagger} &=& \eta^{aa}\stackrel{ab}{(-k)}\,,\quad 
(\stackrel{ab}{(k)})^2 =0\,, \quad \stackrel{ab}{(k)}\stackrel{ab}{(-k)}
=\eta^{aa}\stackrel{ab}{[k]}\,,\nonumber\\
\stackrel{ab}{[k]}^{\dagger} &=& \,\stackrel{ab}{[k]}\,, \quad \quad \quad \quad
(\stackrel{ab}{[k]})^2 = \stackrel{ab}{[k]}\,, 
\quad \stackrel{ab}{[k]}\stackrel{ab}{[-k]}=0\,,
\nonumber\\
\stackrel{ab}{(k)}\stackrel{ab}{[k]}& =& 0\,,\qquad \qquad \qquad 
\stackrel{ab}{[k]}\stackrel{ab}{(k)}=  \stackrel{ab}{(k)}\,, \quad \quad \quad
  \stackrel{ab}{(k)}\stackrel{ab}{[-k]} =  \stackrel{ab}{(k)}\,,
\quad \, \stackrel{ab}{[k]}\stackrel{ab}{(-k)} =0\,,
\nonumber\\
%
\stackrel{ab}{\tilde{(k)}}^{\dagger} &=& \eta^{aa}\stackrel{ab}{\tilde{(-k)}}\,,\quad
(\stackrel{ab}{\tilde{(k)}})^2=0\,, \quad \stackrel{ab}{\tilde{(k)}}\stackrel{ab}{\tilde{(-k)}}
=\eta^{aa}\stackrel{ab}{\tilde{[k]}}\,,\nonumber\\
\stackrel{ab}{\tilde{[k]}}^{\dagger} &=& \,\stackrel{ab}{\tilde{[k]}}\,,
\quad \quad \quad \quad
(\stackrel{ab}{\tilde{[k]}})^2=\stackrel{ab}{\tilde{[k]}}\,,
\quad \stackrel{ab}{\tilde{[k]}}\stackrel{ab}{\tilde{[-k]}}=0\,,\nonumber\\
\stackrel{ab}{\tilde{(k)}}\stackrel{ab}{\tilde{[k]}}& =& 0\,,\qquad \qquad \qquad 
\stackrel{ab}{\tilde{[k]}}\stackrel{ab}{\tilde{(k)}}=  \stackrel{ab}{\tilde{(k)}}\,, 
\quad \quad \quad
  \stackrel{ab}{\tilde{(k)}}\stackrel{ab}{\tilde{[-k]}} =  \stackrel{ab}{\tilde{(k)}}\,,
\quad \, \stackrel{ab}{\tilde{[k]}}\stackrel{ab}{\tilde{(-k)}} =0\,.
\label{graficcliff1}
\end{eqnarray}
One can further find
\begin{eqnarray}
\label{graficfollow1}
S^{ac}\stackrel{ab}{(k)}\stackrel{cd}{(k)} &=& -\frac{i}{2} \eta^{aa} \eta^{cc} 
\stackrel{ab}{[-k]}\stackrel{cd}{[-k]}\,, \quad 
S^{ac}\stackrel{ab}{[k]}\stackrel{cd}{[k]} = 
\frac{i}{2} \stackrel{ab}{(-k)}\stackrel{cd}{(-k)}\,,\nonumber\\
S^{ac}\stackrel{ab}{(k)}\stackrel{cd}{[k]} &=& -\frac{i}{2} \eta^{aa}  
\stackrel{ab}{[-k]}\stackrel{cd}{(-k)}\,, \quad
S^{ac}\stackrel{ab}{[k]}\stackrel{cd}{(k)} = \frac{i}{2} \eta^{cc}  
\stackrel{ab}{(-k)}\stackrel{cd}{[-k]}\,.
\end{eqnarray}

\end{small}

\section{Acknowledgment} 
The author thanks Department of Physics, FMF, University of Ljubljana, Society of Mathematicians, Physicists and Astronomers of Slovenia,  for supporting the research on the {\it spin-charge-family} theory by offering the room and computer facilities and Matja\v z Breskvar of Beyond Semiconductor for donations, in particular for the annual workshops entitled "What comes beyond the standard models", and N.B. Nielsen, L. Bonora and M. Blagojevic for fruitful discussions which have just started on this topic and might hopefully continue.

\end{document}